%% file: main.tex
\pdfoutput=1
\RequirePackage[T1]{fontenc}
\documentclass[12pt]{article}

\usepackage[T1]{fontenc}
\usepackage[utf8]{inputenc}
\usepackage{microtype}

\usepackage{float}
\usepackage{multicol}
\usepackage{bm}
\usepackage{booktabs}
\usepackage{siunitx}
\usepackage{multirow}
\usepackage{amsmath,amssymb,amsfonts,amsthm,bbm}

\usepackage{mathtools}
\usepackage{graphicx}
\usepackage{textcomp}
\usepackage[table]{xcolor}
\usepackage{nicefrac}
\usepackage{braket}
\usepackage{adjustbox}
\usepackage{rotating}
\usepackage[percent]{overpic}
\usepackage{longtable}
\usepackage{algorithm}
\usepackage{algorithmicx}
\usepackage{algpseudocode}
\usepackage{thmtools,thm-restate}
\usepackage{enumitem}
\usepackage{authblk}
\usepackage{subcaption}
\usepackage[pagebackref=true,breaklinks=true,colorlinks,hyperfootnotes=false]{hyperref}
\hypersetup{
  colorlinks,
  citecolor=citeblue,
  linkcolor=firebrick,
  urlcolor=firebrick
  }
\usepackage[nameinlink,capitalize,noabbrev]{cleveref}

\usepackage{fullpage}
\usepackage[sort]{natbib}
\usepackage{orcidlink}
\usepackage{fancyhdr}
\renewcommand{\headrulewidth}{1pt}   %
\newcommand{\runningtitle}{}
\def\BibTeX{{\rm B\kern-.05em{\sc i\kern-.025em b}\kern-.08em
    T\kern-.1667em\lower.7ex\hbox{E}\kern-.125emX}}

\DeclareFontFamily{OT1}{pzc}{}
\DeclareFontShape{OT1}{pzc}{m}{it}{<-> s * [1.10] pzcmi7t}{}
\DeclareMathAlphabet{\pzccal}{OT1}{pzc}{m}{it}

\input{macros.tex}

\title{\LARGE \rmfamily \bfseries
Variational Quantum Attention for Molecular Graph Learning}

\author[1,2\orcidlink{0009-0009-1911-9566}]{Yu-Cheng Lin}
\author[3,4,\orcidlink{0009-0004-7221-3854}]{Yu-Chao Hsu}
\author[4,\orcidlink{0000-0002-1993-1863}]{\authorcr Tai-Yue Li}
\author[4,\orcidlink{0000-0001-8139-6809}]{Nan-Yow Chen}
\author[5,\orcidlink{0000-0003-0114-4826}]{Samuel Yen-Chi Chen}

\affil[1]{Department of Electrophysics, National Yang Ming Chiao Tung University, Hsinchu 300092, Taiwan.}
\affil[2]{Department of Internal Medicine, School of Medicine College of Medicine, Taipei Medical University, Taipei, Taiwan}
\affil[3]{School of Electrical Engineering, Korea Advanced Institute of Science and Technology (KAIST), 291 Daehak-ro, Yuseong-gu, Daejeon 34141, Republic of Korea}
\affil[4]{National Center for High-Performance Computing , National Institutes of Applied Research (NIAR), Hsinchu 300092, Taiwan}
\affil[5]{Brookhaven National Laboratory, Upton NY, USA}

\begin{document}

\maketitle
\begingroup
\renewcommand\thefootnote{}
\endgroup

\clearpage
\begin{abstract}
Molecular property prediction is central to computational drug discovery, where graph neural networks learn to weight neighboring atomic environments during message passing. Yet it remains unclear how variational quantum circuits alter learned attention behavior in molecular graphs. We introduce an edge-aware variational quantum attention mechanism for molecular graph learning, in which the receiving atom, neighboring atom, and connecting bond jointly determine the quantum attention state. Across five molecular property and bioactivity prediction tasks, QGAT achieves competitive performance relative to GATv2, with a consistent improvement on BBBP across all six evaluated circuit ansatzes. We further compare how the quantum and classical attention scores weight molecular structure beyond accuracy. In the Verubecestat BACE1 inhibitor series, QGAT achieves a higher Spearman correlation than GATv2 and assigns positive attributions to several structural changes consistent with reported structure-activity relationships (SARs). This case study shows that the two attention mechanisms can exhibit different prediction and attribution behavior across structurally related BACE1 analogues, while broader validation is required to determine how consistently these differences generalize across chemical series and targets. Circuit ablations further show that performance depends on the circuit design. Together, these results show that variational quantum attention can serve as a viable alternative molecular attention parameterization while inducing circuit- and chemistry-dependent behavior distinct from a matched classical scorer.

\vspace{1em}
\end{abstract}

\section{Introduction}
Molecular property prediction plays a central role in computational drug discovery, where predictive models are used to prioritize compounds before costly experimental evaluation \citep{Berdigaliyev2020}. Molecular graphs provide a natural representation for this task: atoms are represented as nodes, bonds are represented as edges, and molecular properties arise from interactions among local chemical environments. Graph neural networks (GNNs) exploit this graph structure by propagating information along molecular bonds and have been widely applied to physicochemical and bioactivity prediction tasks \citep{pmlr-v70-gilmer17a,Faria2025QGAT}. Graph attention networks (GATs) \citep{GATv22022,Cucurull2017GAT} further introduce learnable attention coefficients that adaptively weight neighboring atoms during message aggregation. Beyond their role in prediction, these learned attention weights characterize how local structural information is differentially weighted during message passing.


We observe that the attention-scoring step in GAT provides a natural entry point for hybrid quantum-classical modeling, since it operates on local atom-pair representations rather than on the entire molecular graph. This local and edge-wise formulation avoids replacing the full node encoder and eliminates the need to encode high-dimensional molecular features directly into a limited quantum state space. By contrast, fully quantum graph models can require quantum resources that grow with graph size; representative formulations encode an $n$-node graph using $n$ qubits \citep{Sebastian2026QGCN, raj2026scalable, Sauvage2026graph}. Such constructions provide a direct quantum representation of graph structure, but substantially increase the cost of simulation on molecular tasks.
For these reasons, we use variational quantum circuits (VQCs)~\citep{abbas2021power,Du2021Learnability,Schuld_data_encoding2021,HsuAdaptive2025} specifically as the attention scorer while retaining the classical pipeline for feature transformation, message aggregation, readout and prediction. This design isolates the effect of the quantum scoring function and allows us to examine how replacing the classical scorer changes molecular prediction and weighting behavior. %


Recent quantum graph attention network (QGAT) models have demonstrated that VQCs can be used to generate attention coefficients from quantum encoded graph representations \citep{QGAT2025}. Existing quantum graph attention formulations are primarily developed for generic graph settings and do not explicitly incorporate molecular bond attributes into the quantum attention computation. This distinction is important for molecular graphs, where the relation between neighboring atoms is chemically defined by the connecting bond rather than by graph adjacency alone \citep{Hehuan2022edge}. Performance alone also does not reveal how replacing a classical attention scorer with VQC changes the weighting of molecular structure during message passing. This leads to two questions: can an edge-aware VQC provide a competitive molecular attention scorer, and does its circuit induced parameterization lead to systematically different molecular weighting behavior?

To address these questions, we formulate an edge-aware molecular variational quantum attention mechanism in which the receiving atom, neighboring atom, and connecting bond jointly determine the attention score. The remaining message passing pipeline is kept unchanged and classical, making the attention scorer the principal architectural intervention between QGAT and its GATv2 counterpart. 

We evaluate the QGAT across five molecular property and bioactivity prediction tasks using scaffold splits, and then examine how replacing the classical attention scorer changes learned molecular weighting. We analyze the resulting behavior using substructure attribution, attention weights, and integrated gradients (IG), and relate these differences to experimentally reported structure-activity relationships (SARs) in the Verubecestat (MK-8931) BACE1 inhibitor series \citep{2016MK8931}.

In summary, our contributions are as follows:
\begin{itemize}
    \item We introduce an edge-aware variational quantum attention mechanism for molecular graphs that jointly incorporates atom and bond information into the quantum attention computation while retaining the surrounding classical message passing pipeline.
    \item We evaluate QGAT across five molecular property and bioactivity tasks under scaffold splits and compare it with a matched GATv2 baseline. Across the evaluated circuit family, QGAT achieves competitive predictive performance relative to matched GATv2, including a consistent improvement on BBBP across all evaluated ansatzes; circuit ablations further reveal strong task dependence on ansatz and variational depth. 
    \item We characterize the resulting prediction behavior using substructure masking, attention analysis, and IG. In the Verubecestat series, QGAT achieves a higher activity ranking correlation than GATv2 and produces several attribution patterns that are directionally consistent with reported SAR.
\end{itemize}

\section{Related Work}
Hybrid quantum--classical machine learning has increasingly been explored for drug-related prediction, typically by using classical neural networks to extract or compress molecular and biological representations before applying parameterized quantum circuits as encoding or prediction modules \citep{Santagati2024Drug, Jing2023}. Representative studies have applied VQC to protein--ligand binding-affinity, drug-response, and molecular property prediction \citep{Domingo2023,Jeong2025, Sagingalieva2023, Min2025, Dong2023}. These studies demonstrate diverse forms of quantum--classical integration, but the quantum component is typically used for representation encoding or downstream prediction rather than for determining graph attention coefficients.

Recent studies in quantum graph learning have incorporated variational quantum circuits (VQCs) into attention-based message passing. \cite{QGAT2025} showed that VQCs can generate multi-head attention coefficients from quantum encoded node representations, while \cite{Faria2025QGAT} used quantum graph encoders to modulate neighboring-node contributions. Our work builds on this line by studying variational quantum attention in molecular graphs, where both atom and bond features are incorporated directly into the attention computation. This raises a question beyond whether quantum attention can match classical performance: does replacing the classical scorer with a VQC change which molecular information the model relies on when making predictions? Answering this question is important because two models with similar accuracy can nevertheless rely on different structural signals, which may affect their behavior on specific chemical series or local structural modifications.

\section{Preliminaries}
We briefly introduce the graph attention network and quantum computing operations used throughout this work.

\subsection{Graph Attention Networks}

GATs \citep{Cucurull2017GAT, GATv22022} aggregate information from neighboring nodes using learned attention coefficients. For a receiving node $u$ and a neighbor $v$, an attention scorer first produces an unnormalized logit $e_{uv}$ from their node representations. The coefficient assigned to $v$ is then obtained by normalizing the logits over the neighborhood $\mathcal{N}_u$:

\begin{equation}
\alpha_{uv}
=
\frac{\exp(e_{uv})}
{\sum_{w\in\mathcal{N}_u}\exp(e_{uw})},
\label{eq:gat_attention}
\end{equation}

The coefficient $\alpha_{uv}$ determines the contribution of node $v$ to the updating of node $u$. Multi-head GATs compute this operation in parallel across multiple heads.

\subsection{Quantum Machine Learning}

Quantum Machine Learning (QML) combines parameterized quantum circuits with classical learning pipelines to construct trainable hybrid models \citep{Cerezo2022QML, Biamonte2017QML, Marcello2019QML}. A typical workflow consists of three stages: a classical transformation that maps the input into a feature representation, a quantum data encoding procedure that prepares the corresponding quantum state, and a VQC whose measurement outcomes are returned to the classical computation.

Let an input be denoted by $\mathbf{x}\in \mathbb{R}^d$. A classical transformation $f_{\boldsymbol{\theta}_c}(\cdot)$, parameterized by $\boldsymbol{\theta}_c$, maps the input to a lower dimensional feature vector

\begin{equation}
\mathbf{z}=f_{\boldsymbol{\theta}_c}(\mathbf{x})\in \mathbb{R}^{n_f}
\label{eq:classical_map}
\end{equation}

To encode $\mathbf{z}$ into a quantum state, we use amplitude encoding. The feature vector is written as $\tilde{\mathbf{z}}\in \mathbb{R}^{2^{n_q}}$ with $\sum_i |\tilde{z}_i|^2=1$. An $n_q$-qubit state is then prepared as 

\begin{equation}
|\psi(\tilde{\mathbf{z}})\rangle
=
V(\tilde{\mathbf{z}})
|0\rangle^{\otimes n_q}
=
\sum_{i=0}^{2^{n_q}-1}
\tilde{z}_i |i\rangle ,
\label{eq:amplitude_encoding}
\end{equation}

where $V(\tilde{\mathbf{z}})$ denotes the state preparation operation. Amplitude encoding represents a $2^{n_q}$-dimensional vector using $n_q$ qubits, reducing the  qubit requirement for high dimensional inputs. However, preparing an arbitrary amplitude encoded state may require substantial circuit depth. 

The encoded state is subsequently processed by a variational circuit $U(\boldsymbol{\theta})$ composed of parameterized single-qubit rotations and entangling gates (e.g. CNOTs). For a circuit with $N_b$ variational blocks,

\begin{equation}
U(\boldsymbol{\theta})
=
\prod_{\ell=1}^{N_b}
U_{\ell}
\bigl(
\boldsymbol{\theta}^{(\ell)}
\bigr)
\label{eq:vqc}
\end{equation}

where the structure of each block $U_{\ell}$ is determined by the chosen circuit ansatz. The resulting quantum state is 

\begin{equation}
|\psi(\tilde{\mathbf{z}},\boldsymbol{\theta})\rangle
=
U(\boldsymbol{\theta})
V(\tilde{\mathbf{z}})
|0\rangle^{\otimes n_q}.
\label{eq:quantum_state}
\end{equation}

Classical information is extracted from the quantum state through expectation values of observables. For an observable operator $\hat{O}_k$, the corresponding readout is: 

\begin{equation}
q_k
=
\left\langle
\psi(\tilde{\mathbf{z}},\boldsymbol{\theta})
\right|
\hat{O}_k
\left|
\psi(\tilde{\mathbf{z}},\boldsymbol{\theta})
\right\rangle 
\label{eq:measurement_general}
\end{equation}

These expectation values provide continuous classical outputs from the quantum circuit and are used in our model to construct the attention logits described in Section~\ref{sec:method}.

\section{Method}
\label{sec:method}
\begin{figure}[t]
    \centering
    \includegraphics[width=\linewidth]{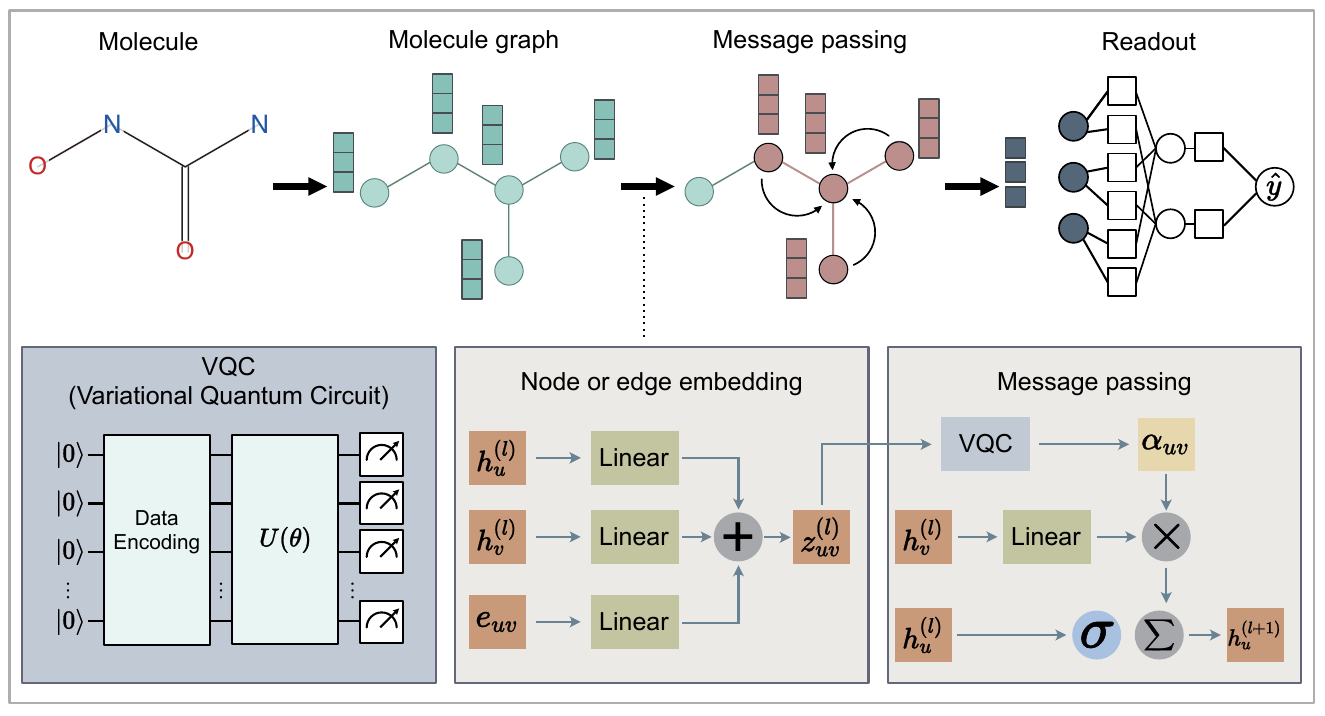}
    \caption{Overview of the edge-aware QGAT architecture for molecular property prediction.}
    \label{fig: workflow}
\end{figure}

We adapt QGAT \citep{QGAT2025} to molecular graphs by incorporating atom and bond representations directly into the quantum attention computation. The quantum module replaces the attention scorer while the message passing and readout prediction remain classical. Figure~\ref{fig: workflow} summarizes the resulting architecture.

\subsection{Molecular graph representation}

Each molecule is represented as a graph $\mathcal{G}=(\mathcal{V}, \mathcal{E})$, where $\mathcal{V}$ denotes the atoms and $\mathcal{E}$ denotes the covalent bonds. Each atom $u$ is associated with a node feature vector $\mathbf{h}_u \in \mathbb{R}^{d}$, and each bond $(u,v)$ is associated with an edge feature vector $\mathbf{e}_{uv}\in \mathbb{R}^{d_e}$. Full details of the molecular graph construction and node and edge descriptors are provided in Appendix ~\ref{app:graph}.

\subsection{Edge-aware variational quantum attention}

For each edge $(u,v)$, the receiving node, neighboring node, and edge features are independently projected into a common multi-head representation. Rather than concatenating the three projected representations, we combine them through element-wise addition:

\begin{equation}
\mathbf{z}_{uv}=\mathbf{W}_u\mathbf{h}_u+ \mathbf{W}_v\mathbf{h}_v + \mathbf{W}_e \mathbf{e}_{uv}\in \mathbb{R}^{Hd_{\mathrm{out}}}
\label{eq:hidden}
\end{equation}

where $H$ is the number of attention heads and $d_{\mathrm{out}}$ is the latent dimension per head. Direct concatenation would increase the latent dimension to $3Hd_{\mathrm{out}}$, requiring either additional qubits or an extra compression projection before quantum encoding. We therefore combine the three projected representations by addition and set the resulting dimension to $Hd_{\mathrm{out}}=2^{n_q}$. The fused representation can then be normalized and directly amplitude encoded without an additional dimensionality reduction layer. In our configurations, $Hd_{\mathrm{out}}$ is 64 or 128, corresponding to 6 or 7 qubits. A direct angle encoding would require 64 or 128 qubits. Therefore, amplitude encoding is particularly relevant because the attention circuit is evaluated for molecular edges throughout message passing.

The encoded state is processed by a variational quantum circuit $U(\boldsymbol{\theta})$. Each variational block consists of trainable single-qubit rotations and entangling operations. We evaluate six circuit ansatzes with different rotation and entangling structures; their complete circuit definitions are provided in Appendix ~\ref{app:ansatz}. 

For each attention head, one Pauli-$Z$ expectation value is used as the corresponding quantum attention logit:

\begin{equation}
e^{(k)}_{uv}=\langle \psi(z_{uv})|U^\dagger(\theta)\hat{O}_k U(\theta)|\psi(z_{uv})\rangle
\label{eq:attention_logit}
\end{equation}

where $\hat{O}_k$ denotes the observable associated with the $k$-th measured qubit. In this work, we choose $\hat{O}_k = Z_k$, i.e., the Pauli-$Z$ operator acting on the $k$-th qubit. We use configurations satisfying $H\le n_q$, assigning each attention head to a distinct Pauli-\(Z\) observable. Because these observables commute and are diagonal in the computational basis, their expectation values can be estimated from the same set of measurement outcomes. 

Since Pauli-$Z$ expectation values are bounded within $[-1,+1]$, the range of pairwise logit differences is also bounded, which limits the sharpness of the subsequent softmax distribution. We therefore introduce a head-specific learnable scale $s_k$: 

\begin{equation}
\tilde{e}^{(k)}_{uv}=s_k e^{(k)}_{uv}
\label{eq:learn_scale}
\end{equation}

The scaled logits are normalized over the incoming neighborhood $u$ using softmax:

\begin{equation}
\alpha^{(k)}_{uv}=
\frac{\exp{\tilde{e}^{(k)}_{uv}}}
{\sum_{w \in N_u} \exp{\tilde{e}^{(k)}_{uw}}}
\label{eq:attention}
\end{equation}

where $\mathcal{N}_u$ denotes the neighbors of receiving node $u$. The resulting attentions are used to weight classical neighbor messages. The output of the $H$ attention heads is concatenated as:

\begin{equation}
\mathbf{h}'_u=
\sigma\bigg(\mathop{\Vert}^H_{k=1} 
\sum_{v\in \mathcal{N}_u} \alpha_{uv}^{(k)} \mathbf{W}^{(k)} \mathbf{h}_v\bigg)
\label{eq:aggregated}
\end{equation}

where $\Vert$ denotes concatenation across attention heads and $\mathbf{W}^{(k)}$ denotes the value projection matrix for the $k$-th head, and $\sigma$ is a non-linear activation function.

\section{Experiments and Results}
\subsection{Experimental Settings}

We evaluate QGAT on five molecular property and bioactivity prediction benchmarks covering both classification and regression tasks. BACE and BBBP are binary classification datasets, while FreeSolv and ESOL are regression benchmarks. We additionally construct a BACE1 bioactivity regression task from ChEMBL \citep{ChEMBL2012}. We focus on molecular property and bioactivity tasks. Large-scale or highly sparse multitask benchmarks, such as HIV, Tox21, SIDER, and MUV are therefore not included in the present study. We use ROC-AUC for classification and RMSE for regression. ChEMBL-BACE additionally reports Spearman rank correlation to measure whether compounds are ordered consistently with their experimental activities, which is particularly relevant for compound prioritization in drug discovery. Dataset statistics and preprocessing details are provided in Appendix ~\ref{apps: dataset}.

To evaluate generalizing beyond closely related molecular structures, we group molecules according to their Bemis--Murcko scaffold splitting \citep{Bemis1996}. Scaffold splitting reduces train--test structural overlap and evaluates generalization to unseen chemical scaffolds than random splitting. We first reserve 10$\%$ of each dataset as a fixed scaffold based test set. The remaining 90$\%$ is partitioned over five folds using scaffold groups, resulting in 72$\%$ training, 18$\%$ validation, and 10$\%$ test data in each fold. Results are reported as mean $\pm$ standard deviation across the five scaffold folds. Further details of the splitting procedure are provided in Appendix ~\ref{app:scaffold}.

To ensure fair comparisons, we adopt consistent experimental settings across all models, including D-MPNN \citep{D-MPNN2019}, AttentiveFP \citep{AttentiveFP2020}, GIN \citep{GIN2018}, GPS \citep{GPS2022}, KA-GAT \citep{li2025kolmogorov}, m-GAT \citep{IGAT2025}, and GATv2 \citep{GATv22022}. GATv2 serves as the primary classical attention baseline because it retains the same graph representation and message passing framework while using a classical attention scorer. All models use the same molecular descriptors, hidden dimension, number of attention heads where applicable, and training settings. Details of the hyperparameter configurations and input feature constructions are provided in Appendix ~\ref{app:hyperparameters}.

\subsection{Main Results on Molecular Property Prediction}

We compare QGAT with seven graph learning baselines across five molecular property prediction tasks. Table~\ref{tab:Tasks} summarizes the strongest observed QGAT configuration for each task together with classical graph-learning baselines. Across tasks, QGAT remains competitive with the matched GATv2 baseline, although the magnitude and direction of the difference depend on the dataset and circuit configuration.

The largest difference between QGAT and GATv2 is observed on BBBP, where the mean ROC-AUC increases from $0.6808$ to $0.7157$. QGAT also achieves higher mean performance on BACE, BBBP, ESOL, and both ChEMBL-BACE metrics, whereas GATv2 obtains a slightly lower RMSE on FreeSolv. These results indicate that the effect of variational quantum attention is task dependent rather than uniformly beneficial across all molecular prediction settings.

Across datasets, QGAT uses approximately $0.5\%$--$1.3\%$ fewer trainable parameters than GATv2. These results show that VQC scorer can replace the classical attention scorer without a systematic loss in performance or an increase in model size. 

\begin{table}[t]
\caption{
Results on molecular property prediction tasks.
Performance is reported as the mean $\pm$ standard deviation over five scaffold folds.
The best and second best results for each metric are highlighted in
\textcolor{red}{red} and \textcolor{blue}{blue}, respectively.
}
\label{tab:Tasks}
\centering

\setlength{\tabcolsep}{3.5pt}
\renewcommand{\arraystretch}{1.1}

\resizebox{\linewidth}{!}{%
\begin{tabular}{lcccccc}
\toprule
\multicolumn{1}{c}{\textbf{Model}}
& \multicolumn{2}{c}{\textbf{Classification}}
& \multicolumn{4}{c}{\textbf{Regression}} \\
\cmidrule(lr){2-3}
\cmidrule(lr){4-7}

& \textbf{BACE}
& \textbf{BBBP}
& \textbf{FreeSolv}
& \textbf{ESOL}
& \multicolumn{2}{c}{\textbf{ChEMBL-BACE}} \\
\cmidrule(lr){6-7}

& \textbf{AUC} $\uparrow$
& \textbf{AUC} $\uparrow$
& \textbf{RMSE} $\downarrow$
& \textbf{RMSE} $\downarrow$
& \textbf{Spearman} $\uparrow$
& \textbf{RMSE} $\downarrow$ \\
\midrule

\textbf{No. of molecules}
& 1,513
& 2,039
& 642
& 1,128
& \multicolumn{2}{c}{4,580} \\

\textbf{Avg. no. of atoms}
& 65
& 46
& 18
& 26
& \multicolumn{2}{c}{59} \\
\midrule

D-MPNN
& $0.7591 \pm 0.0305$
& $0.6713 \pm 0.0230$
& $2.5079 \pm 0.1136$
& $1.0740 \pm 0.0813$
& $0.6520 \pm 0.0266$
& $0.9804 \pm 0.0261$ \\

AttentiveFP
& $0.7872 \pm 0.0337$
& $0.6469 \pm 0.0273$
& $1.9674 \pm 0.3156$
& $0.9672 \pm 0.0666$
& $0.6455 \pm 0.0398$
& $0.9970 \pm 0.0456$ \\

GIN
& $0.7977 \pm 0.0195$
& $0.6670 \pm 0.0207$
& $2.2618 \pm 0.7023$
& \textcolor{blue}{$0.8997 \pm 0.0391$}
& $0.7115 \pm 0.0109$
& $0.9257 \pm 0.0122$ \\

GPS
& $0.7864 \pm 0.0227$
& $0.6580 \pm 0.0414$
& $2.0111 \pm 0.2729$
& \textcolor{red}{$\mathbf{0.8637 \pm 0.0372}$}
& $0.6457 \pm 0.0225$
& $0.9831 \pm 0.0250$ \\

KAGAT
& $0.8005 \pm 0.0231$
& $0.6568 \pm 0.0253$
& $2.3897 \pm 0.2029$
& $1.1366 \pm 0.0451$
& $0.6679 \pm 0.0216$
& $0.9738 \pm 0.0207$ \\

m-GAT
& \textcolor{blue}{$0.8244 \pm 0.0072$}
& \textcolor{blue}{$0.6942 \pm 0.0303$}
& $1.8374 \pm 0.1506$
& $0.9497 \pm 0.0680$
& $0.7166 \pm 0.0294$
& $0.8985 \pm 0.0385$ \\

GATv2
& $0.8210 \pm 0.0202$
& $0.6808 \pm 0.0111$
& \textcolor{red}{$\mathbf{1.7268 \pm 0.2214}$}
& $0.9389 \pm 0.0660$
& \textcolor{blue}{$0.7313 \pm 0.0201$}
& \textcolor{blue}{$0.8895 \pm 0.0330$} \\
\midrule

QGAT
& \textcolor{red}{$\mathbf{0.8283 \pm 0.0133}$}
& \textcolor{red}{$\mathbf{0.7157 \pm 0.0095}$}
& \textcolor{blue}{$1.7400 \pm 0.2033$}
& $0.9103 \pm 0.0317$
& \textcolor{red}{$\mathbf{0.7357 \pm 0.0147}$}
& \textcolor{red}{$\mathbf{0.8800 \pm 0.0188}$} \\
\bottomrule
\end{tabular}%
}

\end{table}

\subsection{Ablation Study: The Effect of Variational Quantum Circuit Design}

We examine two aspects of the variational circuit design: the circuit ansatz and the number of variational layers $N_b$. Rather than searching over unrestricted circuit structures, we consider six shallow circuit ansatzes constructed from parameterized single-qubit gates and CNOT entangling gates. The candidates differ in rotation ordering and CNOT connectivity while keeping the overall circuit scale comparable. A representative circuit block and the complete definitions of all six ansatzes are provided in Appendix~\ref{app:ansatz}.

Table ~\ref{tab:ansatz} compares the six circuit designs across the molecular prediction tasks. No single ansatz consistently dominates across datasets, indicating that circuit architecture is not a neutral implementation detail but a task-dependent modeling choice. 

\begin{table}[h]
\caption{
Performance across six variational circuit ansatzes.
Results are reported as the mean $\pm$ standard deviation across five scaffold folds.
The best result for each evaluation metric is highlighted in \textcolor{red}{red}.
Ansatz~6 results on ChEMBL-BACE are omitted due to the substantially higher computational cost of this configuration under our simulation budget.
}
\label{tab:ansatz}
\centering

\setlength{\tabcolsep}{4pt}
\renewcommand{\arraystretch}{1.1}

\resizebox{\linewidth}{!}{%
\begin{tabular}{lcccccc}
\toprule
\multicolumn{1}{c}{\textbf{Ansatz}}
& \multicolumn{1}{c}{\textbf{BACE}}
& \multicolumn{1}{c}{\textbf{BBBP}}
& \multicolumn{1}{c}{\textbf{FreeSolv}}
& \multicolumn{1}{c}{\textbf{ESOL}}
& \multicolumn{2}{c}{\textbf{ChEMBL-BACE}} \\
\cmidrule(lr){6-7}

&
AUC $\uparrow$
&
AUC $\uparrow$
&
RMSE $\downarrow$
&
RMSE $\downarrow$
&
Spearman $\uparrow$
&
RMSE $\downarrow$ \\
\midrule

1
& \textcolor{red}{$\mathbf{0.8283 \pm 0.0133}$}
& $0.6920 \pm 0.0069$
& \textcolor{red}{$\mathbf{1.7400 \pm 0.2033}$}
& $0.9278 \pm 0.0564$
& $0.7263 \pm 0.0083$
& $0.8955 \pm 0.0140$ \\

2
& $0.8136 \pm 0.0132$
& $0.7023 \pm 0.0092$
& $1.7435 \pm 0.0805$
& $0.9962 \pm 0.1096$
& $0.7228 \pm 0.0152$
& $0.9041 \pm 0.0185$ \\

3
& $0.8143 \pm 0.0131$
& \textcolor{red}{$\mathbf{0.7157 \pm 0.0095}$}
& $1.7942 \pm 0.1943$
& $0.9806 \pm 0.0941$
& $0.7227 \pm 0.0184$
& $0.9041 \pm 0.0261$ \\

4
& $0.8138 \pm 0.0154$
& $0.7000 \pm 0.0148$
& $1.8142 \pm 0.1723$
& $0.9342 \pm 0.0377$
& \textcolor{red}{$\mathbf{0.7357 \pm 0.0147}$}
& \textcolor{red}{$\mathbf{0.8800 \pm 0.0188}$} \\

5
& $0.8063 \pm 0.0105$
& $0.7058 \pm 0.0182$
& $1.8783 \pm 0.2774$
& $0.9288 \pm 0.0559$
& $0.7185 \pm 0.0121$
& $0.9070 \pm 0.0239$ \\

6
& $0.8044 \pm 0.0219$
& $0.6866 \pm 0.0150$
& $1.8838 \pm 0.2699$
& \textcolor{red}{$\mathbf{0.9103 \pm 0.0317}$}
& --
& -- \\
\bottomrule
\end{tabular}%
}

\end{table}

We further vary the number of variational layers over $N_b\in \{1,2,3,4\}$. The complete results are reported in Appendix ~\ref{app:QL}. Increasing circuit depth does not lead to monotonic improvements. BACE and FreeSolv obtain their best performance with a single variational layer, whereas BBBP performs best with three layers and ESOL and ChEMBL-BACE perform best with two layers. Across the evaluated tasks, the strongest configurations are obtained with one to three variational layers, and additional depth does not provide a consistent benefit.

\subsection{Analysis of Molecular Attention and Attribution}
We next compare QGAT and GATv2 at two distinct levels: attention behavior during message passing and attribution with respect to the final prediction. Substructure mask explanation (SME) \citep{Wang2023} measures changes in model output after masking predefined molecular substructures, while integrated gradients (IG) provides signed feature level attributions with respect to the final prediction.  We additionally inspect learned attention weights to characterize how neighboring information is weighted during message passing. 

\textbf{Functional group attribution on ESOL.}
We first apply SME to the ESOL task, where functional group effects admit a comparatively direct physicochemical interpretation. Table ~\ref{tab:SME_ESOL} summarizes representative attribution values, while the complete distributions are reported in Appendix~\ref{app:SME_results}. Both models recover broadly similar attribution directions for several functional groups. Hydrophobic substituents such as $\mathrm{CF_3}$ and $\mathrm{tBu}$ receive negative contributions, whereas several polar groups exhibit positive contributions to predicted solubility. The two models nevertheless differ in attribution magnitude. In particular, QGAT assigns weaker positive contributions than GATv2 to several carbonyl-containing groups, including $\mathrm{C(=O)N}$, $\mathrm{C(=O)CH_3}$, and $\mathrm{C(=O)OMe}$. These results indicate that QGAT preserves several broad structure--solubility trends while exhibiting distinct sensitivities to individual functional groups. 

\begin{table}[h]
\caption{
Representative key functional group attributions produced by GATv2 and QGAT on ESOL.
}
\label{tab:SME_ESOL}
\centering

\setlength{\tabcolsep}{4pt}
\renewcommand{\arraystretch}{1.0}
\begin{tabular}{lccc}
\toprule
\multicolumn{1}{c}{\textbf{Functional groups}}
& \multicolumn{1}{c}{\textbf{GATv2}} 
& \multicolumn{1}{c}{\textbf{QGAT}}\\
\midrule
$\mathrm{-OH}$
& 0.411 & 0.475
 \\
$\mathrm{-NH_2}$
& 0.283 & 0.120
 \\
 
$\mathrm{-CF_3}$ 
& -1.023 & -1.092
 \\

$\mathrm{-tBu}$
& -0.611 & -0.523
 \\

$\mathrm{-C(=O)N}$
& 0.267 & 0.052
 \\

$\mathrm{-C(=O)CH_3}$
& 0.275 & 0.084
 \\

$\mathrm{-C(=O)OMe}$
& 0.496 & 0.185
 \\
\bottomrule
\end{tabular}%

\end{table}

\textbf{Case study on the Verubecestat (MK-8931) BACE1 inhibitor series}. 
We next examine the Verubecestat series reported during BACE1 inhibitor modification \citep{2016MK8931}. No exact molecular matches are present in the ChEMBL-BACE dataset; one analogue has a nearest Tanimoto similarity of 0.92, whereas all remaining compounds have maximum similarity at or below 0.67. We therefore treat this series as an external medicinal chemistry case study rather than a strict chemical out-of-distribution benchmark. This series is particularly informative because it combines a documented medicinal chemistry optimization trajectory with experimentally characterized binding information. Additional results on the Umibecestat \citep{CNP5202021} and LY3202626 \citep{LY3202626} series are reported in Appendix ~\ref{app:additional}.

Across the series, QGAT achieves a higher Spearman correlation than GATv2 ($0.8264$ vs. $0.7954$), whereas GATv2 obtains a slightly lower RMSE ($0.9920$ vs. $1.0306$). Because the ChEMBL-BACE model is trained primarily on standardized IC$_{50}$ values whereas the Verubecestat series is evaluated using $K_i$, we use Spearman correlation as the primary measure of whether the models recover the experimental activity ordering and report RMSE only as a reference.

\begin{figure}[h]
\centering
\includegraphics[width=\linewidth]{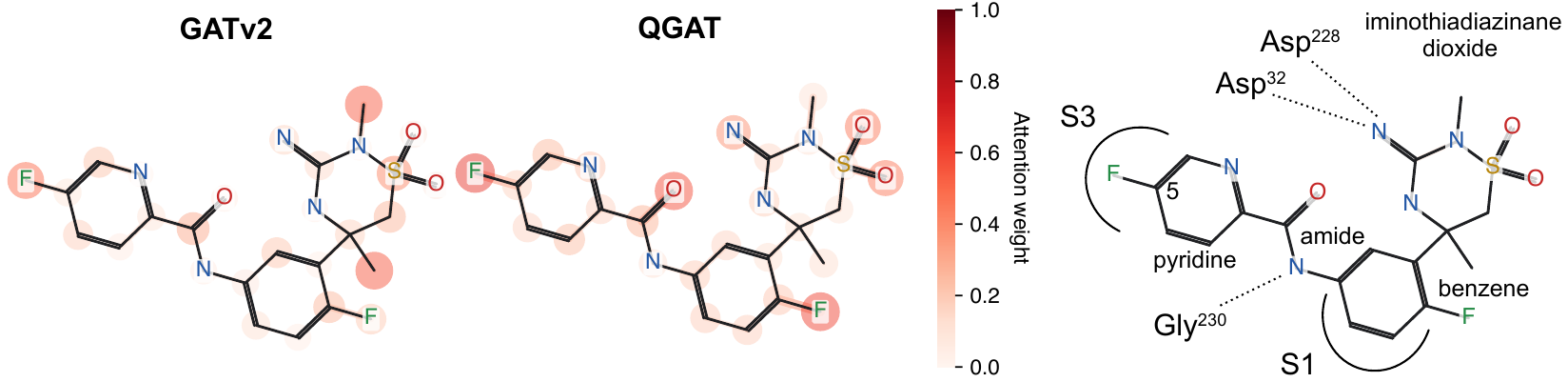}
\caption{Attention visualization for Verubecestat. The right panel highlights the structural groups and binding interactions discussed in the text, with dashed lines indicating hydrogen bonds. QGAT attention scores are transformed for visualization to enhance contrast.}
\label{fig: attention_3}
\end{figure}

The two models also exhibit different attention patterns for Verubecestat (Figure ~\ref{fig: attention_3}). Both assign relatively high attention to the fluoropyridine substituent extending toward the S3 region, while QGAT additionally emphasizes the central amide region, the iminothiadiazinane dioxide scaffold, and the S1-facing fluorophenyl moiety. These regions correspond to structural features involved in the experimentally determined BACE1 binding mode. In the BACE1-Verubecestat cocrystal structure (PDB: 5HU1), the two hydrogen bond donors of the iminothiadiazinane dioxide core interact with the catalytic residues Asp$^{32}$ and Asp$^{228}$, while the amide $\mathrm{NH}$ lies within hydrogen bonding distance of the Gly$^{230}$ carbonyl. The fluoropyridine group occupies the S3 region, with its fluoro substituent extending into the S3 subpocket, whereas the fluorophenyl group projects toward the S1 pocket \citep{2016MK8931}. The QGAT attention pattern overlaps with several ligand regions involved in the experimentally determined binding mode. Because the model uses ligand-derived features only, this correspondence should be interpreted as a qualitative alignment of model weighting with known binding-relevant regions rather than as a reconstruction of protein–ligand interactions.

\begin{figure}[h]
\centering
\includegraphics[width=\linewidth]{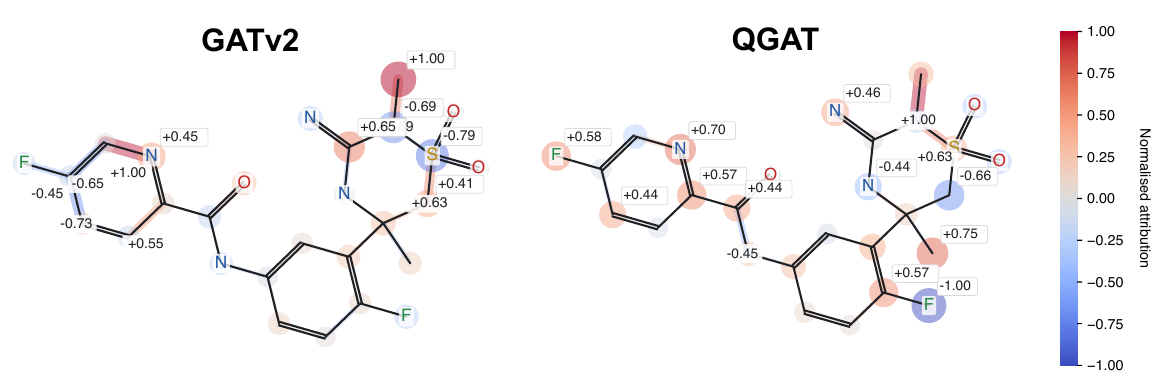}
\caption{Integrated gradients attribution for Verubecestat.}
\label{fig: IG_3}
\end{figure}

Attention weights describe how neighboring information is weighted during message aggregation, but they do not indicate whether a feature contributes positively or negatively to the final prediction. We therefore complement the attention analysis with IG. As shown in Figure ~\ref{fig: IG_3}, QGAT assigns a positive contribution to the 5-fluoropyridine substituent, whereas GATv2 assigns a negative contribution. This direction is consistent with the reported SAR showing that substitution at the pyridyl 5-position improves BACE1 affinity relative to the corresponding unsubstituted analogue \textbf{28} ~\citep{2016MK8931}. Both models assign positive contributions to the pyridine nitrogen, consistent with its role in forming an intramolecular hydrogen bond with the amide $\mathrm{NH}$ that helps lock the preferred picolinamide conformation. 

The attribution analysis also reveals limitations. Both models assign a non-positive contribution to the fluorophenyl fluoride, despite matched analogues demonstrating an approximately fivefold potency improvement from this substitution. Thus, the Verubecestat analysis identifies several structural modifications for which QGAT attribution is directionally consistent with reported SAR and differs from GATv2, while also revealing features that neither model captures correctly.

\textbf{Activity cliff analysis.}
Finally, we examine local structural changes associated with activity cliffs (ACs). We define an AC pair as two structurally similar compounds \citep{AC2014, Ac2019} with a potency difference of $|\Delta pK_i|>1.0$ \citep{ACESGNN2025}. Two pairs in the Verubecestat series satisfy this criterion: compounds \textbf{21}--\textbf{24} and \textbf{30}--\textbf{32}. As shown in Appendix Figures ~\ref{fig:AC_21_24} and ~\ref{fig:AC_30_32}, QGAT exhibits a broader redistribution of IG attribution around the modified pyridyl-amide region for compounds \textbf{21}--\textbf{24} than GATv2. For compounds \textbf{30}-\textbf{32}, both models respond to the benzene-to-pyridine replacement, while QGAT again shows a broader redistribution across the modified region. These observations are qualitative and are used to characterize differences in attribution response rather than to establish a general advantage in AC sensitivity. 

\section{Discussion and Conclusion}
Our results establish edge-aware variational quantum attention as a viable and experimentally competitive mechanism for molecular graph learning. QGAT achieves performance comparable to the matched GATv2 baseline, with the most consistent gain observed on BBBP. Beyond performance, QGAT and GATv2 exhibit distinct molecular weighting and attribution behavior in the Verubecestat series. QGAT achieves a higher activity ranking correlation and assigns attribution directions to several structural modifications that are consistent with reported SAR, while neither model captures all experimentally observed effects. Together, these findings show that replacing the classical scorer with a VQC can alter prediction behavior beyond aggregate accuracy, although they do not establish universally superior chemical interpretability.

The performance of QGAT also depends on the variational circuit design. No single ansatz performs best across all tasks, and increasing the number of variational layers does not consistently improve predictive performance. These results indicate that circuit architecture is an important modeling choice for molecular quantum attention, rather than a design factor that consistently yields performance improvements.

Several limitations remain. First, the present framework is ligand-based and does not explicitly incorporate protein or binding site information, limiting its ability to model interaction specific affinity prediction. Second, the quantum circuits are evaluated using classical simulation rather than real quantum hardware. Third, the current chemical interpretation is based primarily on qualitative attribution analyses and a single case study. The present evidence therefore does not establish generally superior chemical interpretability, and broader quantitative evaluation is needed. Future work should explore protein-aware extensions and quantitative protocols for evaluating chemically relevant attention and attribution.

\section*{Acknowledgment}
Y.-C. Lin and Y.-C. Hsu thank the National Center
for High-Performance Computing (NCHC), National Institutes of Applied Research (NIAR), Taiwan, for providing computational and storage resources supported by the National Science and Technology Council (NSTC), Taiwan, under Grants No.
NSTC 114-2119-M-007-013. 

\bibliographystyle{iclr2027_conference}
\bibliography{iclr2027_conference}
\clearpage

\appendix
\renewcommand{\thesection}{\Alph{section}}
\renewcommand{\thesubsection}{\thesection.\arabic{subsection}}
\addtocontents{toc}{\protect\setcounter{tocdepth}{2}}
\begin{center}
{\LARGE\bfseries Appendix}
\end{center}
\vspace{1.5em}

\renewcommand{\contentsname}{Table of Contents}
\tableofcontents

\clearpage

\section{Experimental Details}
\label{app:experiment}

\subsection{Datasets and Preprocessing}
\label{apps: dataset}
We evaluate the models on five molecular property and bioactivity prediction datasets spanning classification and regression tasks. Table~\ref{tab:dataset} summarizes the dataset size, prediction tasks, and target properties used in this study.

\textbf{BACE}
BACE \citep{BACE2016} dataset measures the binding results for a set of inhibitors of $\beta$-secretase 1 (BACE-1) and is gathered in MoleculeNet \citep{MoleculeNet2018}.

\textbf{BBBP}
The Blood-Brain Barrier Penetration (BBBP) \citep{BBBP2012} is a binary classification benchmark containing measurements of whether a small molecule can penetrate the central nervous system. 

\textbf{ESOL}
The ESOL \citep{ESOL2004} dataset contains aqueous solubility values (solubility in mols per liter) for small organic molecules in log scale.

\textbf{FreeSolv}
FreeSolv \citep{FreeSolv2014} contains experimental and calculated hydration free energy values for small molecules in water.

\textbf{ChEMBL-BACE}  
We construct a BACE1 bioactivity regression dataset from ChEMBL \citep{ChEMBL2012}. We retain measurements associated with a single human BACE1 target and use reported IC$_{50}$ values.  IC$_{50}$ values are transformed to pIC$_{50}$. The resulting activity values are subsequently normalized for model training.

\begin{table}[h]
\caption{
Summary for the molecular property and bioactivity datasets used in this study
}
\label{tab:dataset}
\centering

\setlength{\tabcolsep}{4pt}
\renewcommand{\arraystretch}{1.1}

\resizebox{\linewidth}{!}{%
\begin{tabular}{lcccccc}
\toprule
\multicolumn{1}{c}{Dataset}
& \multicolumn{1}{c}{Task}
& \multicolumn{1}{c}{No. Tasks}
& \multicolumn{1}{c}{Molecules}
& \multicolumn{1}{c}{Target} \\
\midrule

BACE
& Classification & 1 & 1,513 &
Binding results for a set of inhibitors of human BACE1
\\

BBBP
& Classification & 1 & 2,039 &
Permeability properties of the BBB penetration
\\

FreeSolv
& Regression & 1 & 642 & 
Hydration free energy
\\

ESOL
& Regression & 1 & 1,128 &
Water solubility 
\\

ChEMBL-BACE
& Regression & 1 & 4,580 & 
Binding results for a set of inhibitors of human BACE1
\\
\bottomrule
\end{tabular}%
}

\end{table}

\subsection{Scaffold Splitting}
\label{app:scaffold}
A molecular scaffold describes the core structural framework of a molecule. In the Bemis--Murcko scaffolds \citep{Bemis1996}, the scaffold is composed primarily of ring systems and the linkers connecting them, while terminal side chains are removed. Consequently, molecules that differ in peripheral substituents can still share the same scaffold if they are built around the same ring-and-linker framework. 

Scaffolds are particularly relevant in medicinal chemistry because compounds within the same chemical series are often generated by modifying substituents around a conserved core \citep{Ye2016scaffold}. As a result, structurally related analogues can share both a scaffold and similar physicochemical or biological properties. Under a random split, close analogues may therefore appear in both the training and test set, allowing a model to be evaluated on molecules that are structurally similar to those observed during training \citep{D-MPNN2019}.

To provide a more stringent evaluation of structural generalization, we use scaffold splitting. Each molecule is first assigned a Bemis--Murcko scaffold using RDKit \citep{rdkit2016}, and molecules sharing the same scaffold are grouped together. Scaffold groups are then assigned to the training, validation, and test partitions without splitting molecules from the same scaffold across partitions \citep{MoleculeNet2018, D-MPNN2019}. Thus, the evaluation emphasizes generalization to molecular cores that are not represented in the corresponding training set \citep{MoleculeNet2018}.

In our experiments, we first reserve 10\% of the molecules as a fixed scaffold based test set. The remaining molecules are partitioned into five folds, producing an 72\% training and 18 \% validation set for each fold. 

\begin{figure}[h]
\centering
\includegraphics[width=\textwidth]{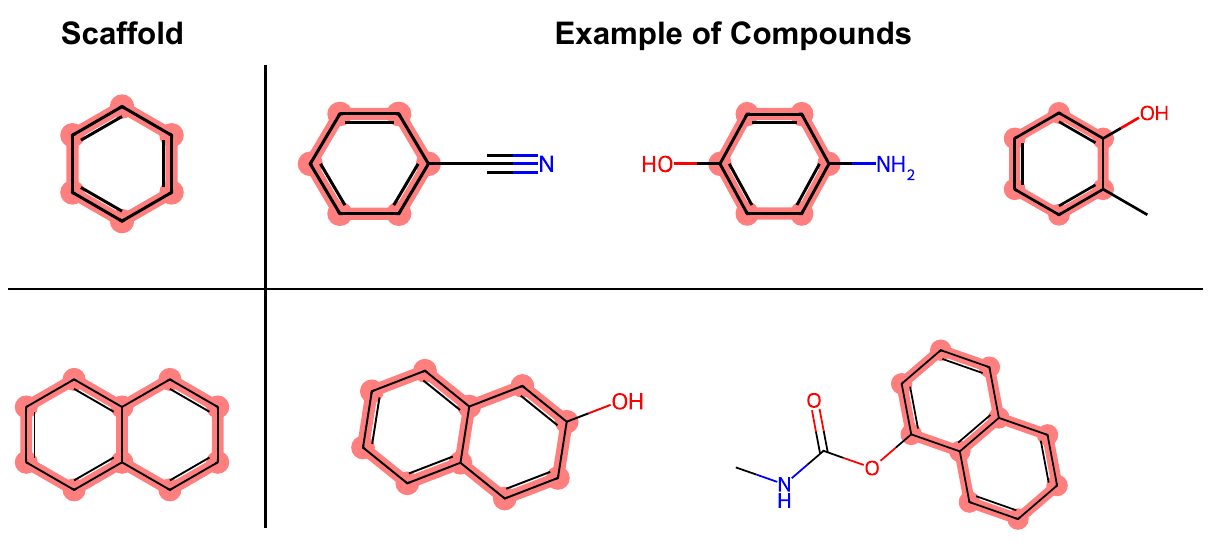}
\caption{Examples of molecular scaffolds and compounds sharing the corresponding core structures. Highlighted atoms indicate the Bemis--Murcko scaffold, while peripheral substituents vary among compounds within the same scaffold group.}
\label{fig: scaffold}
\end{figure}

\subsection{Graph Representation Method}
\label{app:graph}

We represent each molecule as a bidirectional graph $G=(V,E)$, where atoms, including explicit hydrogen atoms, correspond to nodes ($V$) and each covalent bond is represented by two parallel edge types ($E$), namely a chemical bond edge and a charge--distance edge. Each atom is encoded by a 92-dimensional vector following the atom representation used in crystal graph convolutional neural network \citep{CGNN2018}, incorporating atomic number, group, period, electronegativity and covalent radius. Both edge types are represented by 21-dimensional feature vectors. For a chemical bond edge, the first 15 dimensions encode the direction, bond type, bond distance, and ring membership, while the remaining six dimensions are set to zero. For a charge--distance edge, the first 15 dimensions are set to zero, whereas the final six dimensions contain the partial charges $q_i$ and $q_j$, their pairwise product $q_iq_j$, and the inverse-distance features $r_{ij}^{-1}$, $r_{ij}^{-6}$, and $r_{ij}^{-12}$. The charge and inverse distance terms provide information about the local electronic environment and three-dimensional geometry of the bonded atom pair. The inverse-distance terms are used as nonlinear geometric descriptors rather than as explicit estimates of interaction energies. Detailed descriptions of the feature vectors are summarized in Table ~\ref{tab:graph}.

\begin{table}[h]
\caption{Feature Types used for molecular graph construction}
\label{tab:graph}
\begin{center}
\resizebox{\columnwidth}{!}{
\begin{tabular}{lllcc}
\multicolumn{1}{c}{Entity} &
\multicolumn{1}{c}{Feature Type} &
\multicolumn{1}{c}{Description} &
\multicolumn{1}{c}{Type} &
\multicolumn{1}{c}{Size} \\
\midrule
Atom &
CGCNN &
Atomic number, radius, and electronegativity &
One-Hot & 92 \\
\midrule
Chemical bond edge
& Bond direction & None, beginwedge, begindash, etc. & One-Hot & 7 \\
& Bond type & Single, Double, Triple, or Aromatic & One-Hot & 4 \\
& Bond length & $r_{ij}$ and $r_{ij}^2$ & Float & 2 \\
& Ring membership & Whether the bond belongs to a ring & One-Hot & 2 \\
\midrule
Charge-distance edge
& Partial charges & Atomic charges in molecule & Float & 3 \\
& Distance between atoms & Distance terms $(r_{ij}^{-1}, r_{ij}^{-6}, r_{ij}^{-12})$ & Float & 3 \\
\bottomrule
\end{tabular}
}
\end{center}
\end{table}

\subsection{Hyperparameter and Training Details}
\label{app:hyperparameters}
All models were implemented in PyTorch \citep{paszke2017automatic} and trained on NVIDIA H200 GPUs. Unless otherwise specified, we use the Adam optimizer with a learning rate of $1\times 10^{-3}$ with weight decay $1\times 10^{-4}$, and dropout rate of $0.05$. All models are trained for a maximum of 500 epochs. For classification tasks, we use binary cross-entropy loss, whereas regression tasks are optimized using mean squared error loss. Validation performance is used only for checkpoint selection within each trained configuration. The implementation will be released publicly upon publication.

Table ~\ref{tab:hyperparameters} summarizes the model configurations used for each dataset. The hidden dimension, number of attention heads, number of layers, learning rate, weight decay, dropout rate, maximum training epochs, and batch size are kept identical between QGAT and the corresponding GATv2 baseline for each task. The quantum module introduces computational and memory overhead during classical circuit simulation, particularly because attention scores must be evaluated for molecular edges and multiple attention heads. We therefore restrict the search space to relatively compact hidden dimensions and four attention heads rather than performing an exhaustive architecture search over substantially larger models. 

\begin{table}[h]
\caption{GAT and QGAT hyperparameters for different datasets.}
\label{tab:hyperparameters}
\begin{center}
\resizebox{\columnwidth}{!}{
\begin{tabular}{llcccccccc}
\multicolumn{1}{c}{\bf Dataset} &
\multicolumn{1}{c}{\bf Model} &
\multicolumn{1}{c}{\bf Ansatz} &
\multicolumn{1}{c}{\bf Hidden} &
\multicolumn{1}{c}{\bf Heads} &
\multicolumn{1}{c}{\bf Layers} &
\multicolumn{1}{c}{\bf $N_b$} &
\multicolumn{1}{c}{\bf Qubits} &
\multicolumn{1}{c}{\bf Batch} &
\multicolumn{1}{c}{\bf Params.}
\\ \hline \\
BACE        & QGAT & 1 & 16 & 4 & 2 & 1 & 6 & 64  & 13,153 \\
BACE        & GATv2  & -- & 16 & 4 & 2 & -- & -- & 64  & 13,249 \\
BBBP        & QGAT & 3 & 32 & 4 & 4 & 3 & 7 & 64  & 73,385 \\
BBBP        & GATv2  & -- & 32 & 4 & 4 & -- & -- & 64  & 73,729 \\
FreeSolv    & QGAT & 1 & 32 & 4 & 4 & 1 & 7 & 64  & 73,273 \\
FreeSolv    & GATv2  & -- & 32 & 4 & 4 & -- & -- & 64  & 73,729 \\
ESOL        & QGAT & 6 & 16 & 4 & 2 & 2 & 6 & 64  & 13,201 \\
ESOL        & GATv2  & -- & 16 & 4 & 2 & -- & -- & 64  & 13,377 \\
ChEMBL BACE & QGAT & 4 & 32 & 4 & 4 & 2 & 7 & 256 & 73,345 \\
ChEMBL BACE & GATv2  & -- & 32 & 4 & 4 & -- & -- & 256 & 73,729 \\
\end{tabular}
}
\end{center}
\end{table}

Unless otherwise stated, quantum expectation values during training are computed using exact state-vector simulation. Noisy quantum simulations are treated separately in the corresponding robustness experiments rather than during the standard training procedure.

\section{Variational Quantum Circuit Details}
\label{app:VQC}

\subsection{Circuit Architectures}
\label{app:ansatz}

We consider a constrained family of shallow parameterized quantum circuits. Following the hardware-efficient circuit family studied by \cite{Sukin2019}, we construct each ansatz from parameterized single-qubit rotation gates and CNOT entangling gates.

For a rotation around axis $a\in \{x,y,z\}$, the corresponding single-qubit gate is

\begin{equation}
R_a(\theta)= \exp 
(-i\frac{\theta}{2}\sigma_a)
\label{eq:rotation_gate}
\end{equation}

where $\sigma_a$ denotes the Pauli matrix associated with the rotation
axis.

Because rotations about different axes generally do not commute, their ordering changes the resulting variational unitary. For example,

\begin{equation}
R_Y(\theta_1)R_Z(\theta_2)\neq
R_Z(\theta_2)R_Y(\theta_1)
\label{eq:rotation_commute}
\end{equation}

\begin{figure}[h]
\centering
\includegraphics[width=\textwidth]{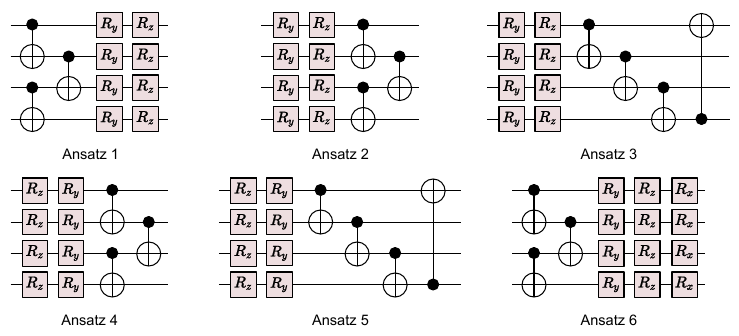}
\caption{Six ansatzes evaluated in this study. For illustration, the circuits are shown with $n_q=4$ qubits and $N_b=1$ block. All ansatzes use  shallow hardware-efficient circuits composed of parameterized single-qubit rotations and CNOT entangling gates, while differing in rotation patterns and entangling connectivity.}
\label{fig: ansatz}
\end{figure}

Accordingly, changing the rotation sequence modifies the local parameterization even when the circuit uses the same gate set and a similar number of trainable parameters. Figure~\ref{fig: ansatz} summarizes the six circuit templates considered in our experiments. The candidates share the same general design principle of shallow single-qubit rotations followed by CNOT gates, while varying the rotation sequence and entangling connectivity. Previous studies have shown that the choice of parameterized gates, entangling structure, and circuit depth can substantially affect properties such as expressibility and entangling capability \citep{Sukin2019}; however, these circuit structures do not by themselves determine predictive performance for a given learning task. We therefore evaluate the six candidates directly on the molecular prediction tasks rather than assuming a universally preferred circuit structure.  As reported in Table~\ref{tab:ansatz}, no single ansatz performs best across all datasets, indicating that performance is
sensitive to the circuit parameterization and task.

\subsection{Effect of Variational Depth}
\label{app:QL}

We examine whether increasing the depth of the VQC consistently improves performance. Table~\ref{tab:QL} reports the results obtained by varying the number of variational layers while keeping the remaining model hyperparameters fixed for each task.

The optimal circuit depth is task-dependent, and increasing the number of blocks does not lead to performance improvements. BACE and FreeSolv achieve their best performance with a single block. In contrast, ESOL and ChEMBL-BACE perform best with two blocks, while BBBP reaches its highest ROC-AUC with three blocks. 

Deeper circuits generally provide limited additional benefits beyond these task-specific optima. For example, BACE performance decreases from $0.8283$ with one block to approximately $0.80$ with two and three blocks, while FreeSolv RMSE gradually increases as the circuit depth grows. A similar degradation is observed for BBBP and ESOL at four blocks after reaching their respective optima at shallower depths.

These results suggest that the benefit of additional circuit depth is not universal. Relatively shallow circuits are sufficient for the evaluated molecular property prediction tasks, whereas increasing circuit depth may introduce additional optimization complexity without providing a corresponding improvement in performance.

\begin{table}[h]
\caption{
Effect of variational circuit depth on QGAT performance.
Results are reported as mean $\pm$ standard deviation across five folds.
The best result for each task is shown in {\color{red}{red}}.
Unreported results are denoted by ``--'' due to higher computational cost.
}
\label{tab:QL}
\centering

\setlength{\tabcolsep}{4pt}
\renewcommand{\arraystretch}{1.1}

\resizebox{\linewidth}{!}{%
\begin{tabular}{lcccccc}
\toprule
\multicolumn{1}{c}{$\boldsymbol{N_b}$}
& \multicolumn{1}{c}{\textbf{BACE}}
& \multicolumn{1}{c}{\textbf{BBBP}}
& \multicolumn{1}{c}{\textbf{FreeSolv}}
& \multicolumn{1}{c}{\textbf{ESOL}}
& \multicolumn{2}{c}{\textbf{ChEMBL-BACE}} \\
\cmidrule(lr){6-7}

&
AUC $\uparrow$
&
AUC $\uparrow$
&
RMSE $\downarrow$
&
RMSE $\downarrow$
&
Spearman $\uparrow$
&
RMSE $\downarrow$ \\
\midrule

1
& \textcolor{red}{$\mathbf{0.8283 \pm 0.0133}$}
& $0.7019 \pm 0.0164$
& \textcolor{red}{$\mathbf{1.7400 \pm 0.2033}$}
& $0.9263 \pm 0.0275$
& $0.7265 \pm 0.0198$

& $0.9018 \pm 0.0130$ \\

2
& $0.8023 \pm 0.0235$
& $0.7125 \pm 0.0095$
& $1.7697 \pm 0.1798$
& \textcolor{red}{$\mathbf{0.9103 \pm 0.0317}$}
& \textcolor{red}{$\mathbf{0.7357 \pm 0.0147}$}
& \textcolor{red}{$\mathbf{0.8800 \pm 0.0188}$} \\

3
& $0.8018 \pm 0.0126$
& \textcolor{red}{$\mathbf{0.7157 \pm 0.0095}$}
& $1.8498 \pm 0.2731$
& $0.9218 \pm 0.0538$
& --
& -- \\

4
& $0.8119 \pm 0.0118$
& $0.6919 \pm 0.0153$
& $1.8816 \pm 0.2340$
& $0.9377 \pm 0.0645$
& --
& -- \\
\bottomrule
\end{tabular}%
}

\end{table}

\subsection{Post-training Robustness to Simulated Quantum Noise}
\label{app:noise}

We evaluate the post-training sensitivity of QGAT to two simulated noise channels: depolarizing noise and amplitude damping. All models are trained under noiseless simulation and subsequently evaluated with noise injected into the VQC during inference. 

\textbf{Depolarizing noise.}
For a single-qubit gate acting on qubit $q$, we apply the depolarizing channel \citep{Nielsen2010}

\begin{equation}
\mathcal{E}_{\mathrm{dep},1Q}(\rho)
=
(1-p_{1Q})\rho
+
\frac{p_{1Q}}{3}
\sum_{P\in \{X,Y,Z\}}
P_q \rho P_q
\label{eq:dep_1q}
\end{equation}

where $p_{1Q}$ denotes the total probability of a non-identity single-qubit Pauli probability. 
For a CNOT acting on control qubit $c$ and target qubit $t$, we use a joint two-qubit depolarizing channel:

\begin{equation}
\mathcal{E}_{\mathrm{dep},2Q}^{(c,t)}(\rho)
=
(1-p_{2Q})\rho
+
\frac{p_{2Q}}{15}
\sum_{P\in\mathcal{P}_2\setminus\{I\otimes I\}}
P\rho P^\dagger,
\label{eq:dep_2q}
\end{equation}

where $\mathcal{P}_2=\{I,X,Y,Z\}^{\otimes 2}$ and $p_{2Q}$ denotes the total probability assigned to the 15 non-identity two-qubit Pauli operators. 

\textbf{Amplitude damping.}

Amplitude damping describes irreversible relaxation from the excited state $|1\rangle$ toward the ground state $|0\rangle$. For a single qubit, the channel is defined as \citep{Nielsen2010, Wang2025amplitude}

\begin{equation}
\mathcal{E}_{\mathrm{AD}}(\rho)= K_0\rho K_0^\dagger + K_1\rho K_1^\dagger
\label{eq:ad_1q}
\end{equation}

with Kraus operators

\begin{equation}
K_0 = 
\begin{bmatrix}
1 & 0 \\
0 & \sqrt{1-p_{\mathrm{AD}}}
\end{bmatrix}, 
\qquad
K_1 = 
\begin{bmatrix}
0 & \sqrt{p_{\mathrm{AD}}} \\
0 & 0
\end{bmatrix}
\label{eq:ad_kraus}
\end{equation}

where $p_{\mathrm{AD}}\in [0,1]$ denotes the damping probability. After each single-qubit rotation, the amplitude damping channel is applied to the corresponding qubit. 

For a CNOT acting on qubits $c$ and $t$, amplitude damping is applied independently to both participating qubits using the same damping strength $p_{\mathrm{AD}}$:

\begin{equation}
\mathcal{E}^{(c,t)}_{p_{\mathrm{AD}}}(\rho)
=
\sum_{i,j\in\{0,1\}}
(K_i^{(c)}\otimes K_j^{(t)})
\rho
(K_i^{(c)}\otimes K_j^{(t)})^\dagger.
\label{eq:ad_2q}
\end{equation}

Equivalently, this local two-qubit channel can be represented by the four product Kraus operators $M_{ij}=K_i\otimes K_j$, with $i,j\in\{0,1\}$.  This model assumes independent relaxation of the two qubits and does not include correlated decay.

\textbf{Noisy circuit simulation.}
Amplitude encoding is performed without noise and initially prepares a state vector. For noisy inference, the state is converted to the density matrix. Noise is applied after every gate. Specifically, each gate rotation is followed by the corresponding single-qubit noise channel, whereas each CNOT is followed by either the joint two-qubit depolarizing channel in Eq.~\ref{eq:dep_2q} or amplitude-damping channels in Eq.~\ref{eq:ad_2q}. The resulting evolution can be summarized as

\begin{equation}
\rho
\xrightarrow{U}
U\rho U^\dagger
\xrightarrow{\mathcal{E}}
\mathcal{E}(U\rho U^\dagger).
\end{equation}

The noise channels are evaluated directly through density matrix evolution by summing the corresponding Pauli or Kraus branches. 

\textbf{Evaluation protocol.}
 We evaluate the channel strengths $p\in\{0.001, 0.01,0.1,0.2\}$ where $p$ denotes $p_{\mathrm{dep}}$ for depolarizing noise and $p_{\mathrm{AD}}$ for amplitude damping. For the reported experiments, the same strength is assigned to single- and two-qubit operations within each noise model.

We focus on post-training inference sensitivity rather than training under noisy conditions. Noisy training would require differentiating through density matrix evolution after each quantum operation, substantially increasing the computational and memory cost relative to the noiseless state vector training used in the main experiments. 

\begin{figure}[t]
\centering
\includegraphics[width=\linewidth]{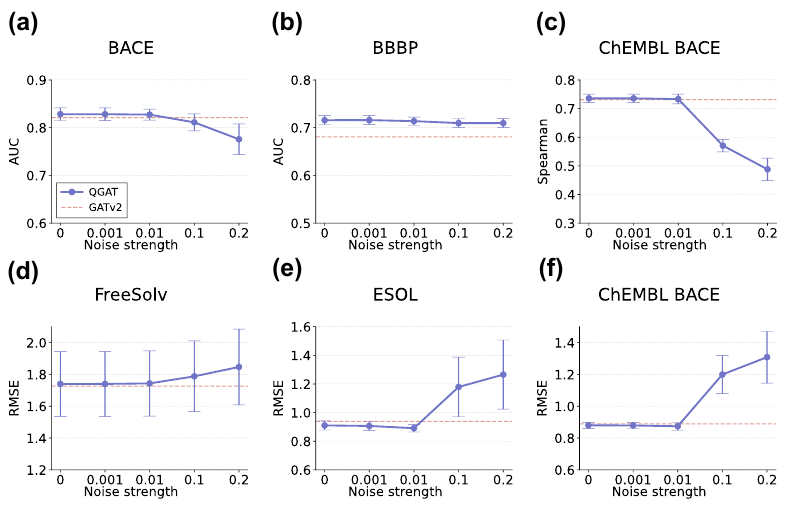}
\caption{Performance of QGAT at inference stage under depolarizing noise. Models are trained under noiseless simulation and evaluated with depolarizing noise injecting into the VQC at different channel strengths. Error bars indicate variation across five scaffold folds.}
\label{fig: depolarizing}
\end{figure}

\begin{figure}[h]
\centering
\includegraphics[width=\linewidth]{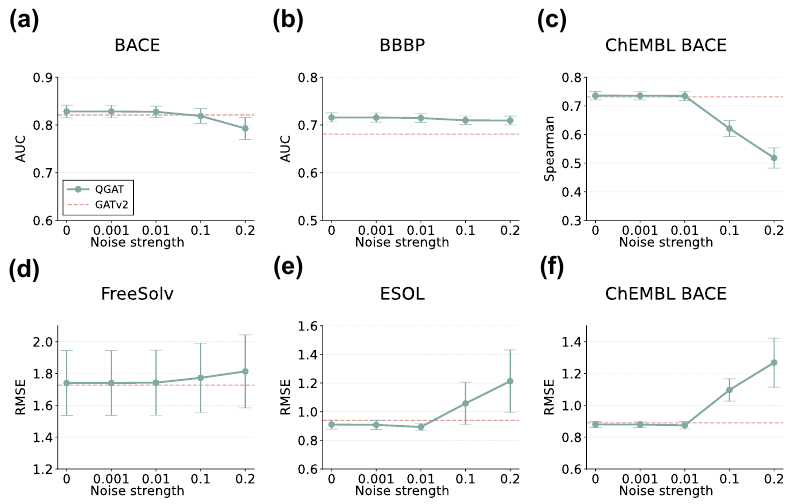}
\caption{Performance of QGAT at inference stage under amplitude damping noise. Models are trained under noiseless simulation and evaluated with amplitude damping injecting into the VQC at different channel strengths. Error bars indicate variation across five scaffold folds.}
\label{fig: amplitude}
\end{figure}

As shown in Figures ~\ref{fig: depolarizing} and ~\ref{fig: amplitude}, QGAT remains largely stable under weak noise. At channel strengths up to $0.01$, the performance remains close to the noiseless setting across all tasks. In contrast, at larger noise strengths, the degradation becomes task dependent.  BACE and FreeSolv exhibit gradual performance deterioration as the noise strength increases, whereas ESOL and ChEMBL-BACE show a substantially sharper degradation at $p\geq 0.1$. Amplitude damping generally produces slightly smaller performance losses than depolarizing noise at the larger error rates.

BBBP is comparatively insensitive to both noise models, its ROC-AUC changes only marginally even at the largest noise strength. This observation suggests that the prediction on BBBP is relatively stable under perturbations to the circuit, although the mechanism underlying this behavior remains unclear.

Under matched numerical channel parameters, amplitude damping generally produces smaller performance changes than depolarizing noise at the larger tested values. Overall, QGAT shows limited performance degradation under weak simulated noise, whereas stronger noise produces larger and task-dependent degradation.

\section{Substructure Attribution Analysis}
\label{app:SME}

\subsection{Substructure Mask Explanation}
\label{app:SME_method}

Substructure Mask Explanation (SME) \citep{Wang2023} is a perturbation-based interpretation method that estimates the contribution of a molecular substructure by comparing the model prediction before and after masking that substructure. By aggregating the attribution scores over functional groups occurrences across the entire dataset, we examine how specific chemical groups influence the predictions of each model. 

Following the original SME formulation, masking is performed after message passing rather than by deleting atoms or bonds from the input graph.  The complete molecular graph is first processed by the GNN to obtain the final node representations, after which the contributions of nodes belonging to the selected substructure are suppressed during graph readout.  We adapt this procedure to the pooling operators used by our models.

\begin{equation}
\{h_v\}_{v=1}^{N}=
f_{\theta}^{\mathrm{GNN}}(G)
\label{eq:sme_node_embedding}
\end{equation}

where $h_v$ is the learned representation of atom $v$ after the final message passing layers. The graph structure and message passing process are therefore kept unchanged, and masking is applied only at the readout layer.

For a functional-group occurrence $S\subseteq V$, we define the binary node mask

\begin{equation}
m_v^{(S)}=
\begin{cases}
0, & v\in S, \\
1, & v\notin S
\end{cases}
\label{eq:sem_mask}
\end{equation}

Let $\mathcal{R}$ denote the pooling used by the corresponding model configuration. The unmasked molecular representation is:

\begin{equation}
g
=
\mathcal{R}\left(\{h_v\}_{v\in V}\right)
\label{eq:sme_unmasked_pooling}
\end{equation}

The corresponding masked molecular representation is 

\begin{equation}
g^{(S)}
=
\mathcal{R}\left(\{m_v^{(S)}h_v\}_{v\in V};\,S\right)
\label{eq:sme_masked_pooling}
\end{equation}

where $\mathcal{R}_{S}$ applies the same readout rule while suppressing the contribution of nodes in $S$.  For mean pooling, masked node embeddings are set to zero while the denominator remains the original number of nodes $N$.  For max pooling, masked nodes are excluded from the element-wise maximum.  Depending on the model configuration, the mean and max pooled representations are either concatenated or combined by element-wise addition.

The original and masked predictions are then computed using the same prediction head:

\begin{equation}
\hat{y}
=
f_{\theta}^{\mathrm{head}}(g),
\qquad
\hat{y}^{(S)}
=
f_{\theta}^{\mathrm{head}}\left(g^{(S)}\right).
\label{eq:sme_predictions}
\end{equation}

The attribution of substructure occurrence $S$ is defined as 

\begin{equation}
A_S
=
\hat{y}-\hat{y}^{(S)}.
\label{eq:sme_attribution}
\end{equation}

A positive attribution indicates that masking the substructure decreases the model output, whereas a negative attribution indicates that masking the substructure increases it.  The sign therefore reflects the direction of the model output and should not be interpreted as an intrinsically favorable or unfavorable chemical effect.

\subsection{Complete Attribution Distributions}
\label{app:SME_results}

To complement the representative functional group analysis reported in the main text, we provide the complete SME attribution distributions for all evaluated datasets. For each functional group, the distribution contains attribution scores from all matched occurrences in the corresponding dataset, thereby retaining variation across different molecular contexts rather than reporting only the dataset-level mean.

\begin{figure}[h]
\begin{center}
\includegraphics[width=\linewidth]{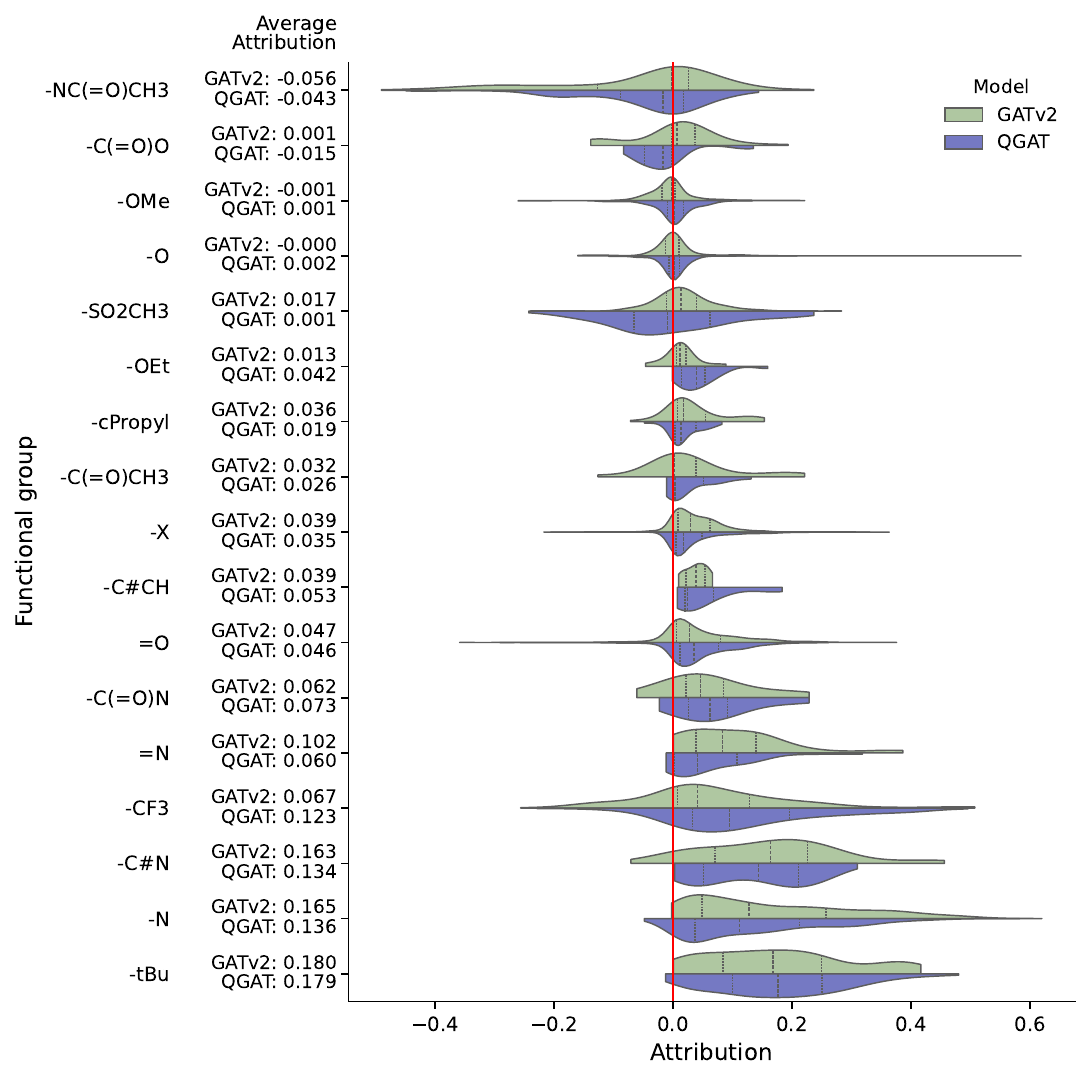}
\end{center}
\caption{The attributions of different functional groups in the whole BACE dataset. Green colors represent GATv2 model, while blue colors represent QGAT model. }
\label{fig: SME_BACE}
\end{figure}

\begin{figure}[h]
\begin{center}
\includegraphics[width=\linewidth]{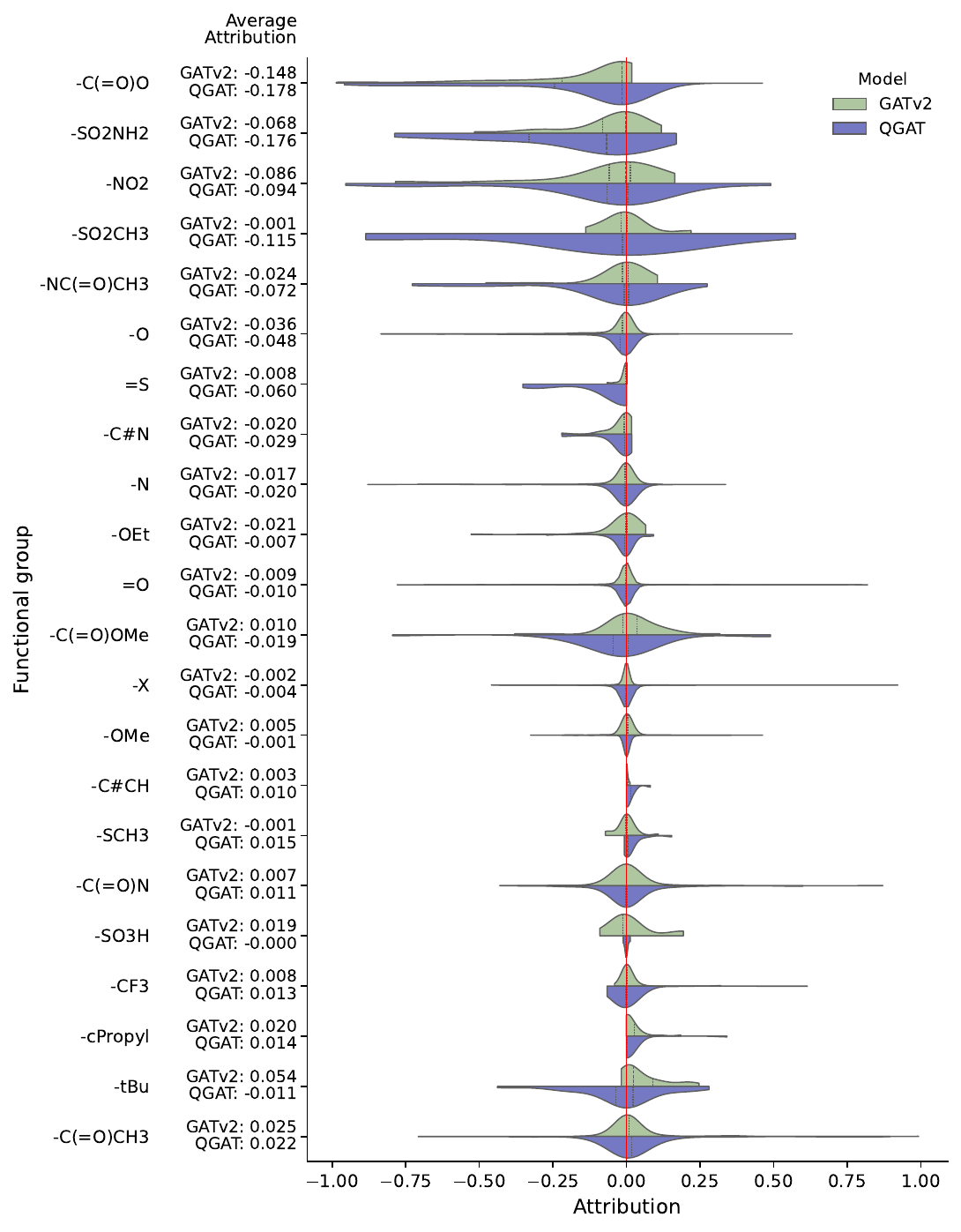}
\end{center}
\caption{The attributions of different functional groups in the whole BBBP dataset. Green colors represent GATv2 model, while blue colors represent QGAT model. }
\label{fig: SME_BBBP}
\end{figure}

\begin{figure}[h]
\begin{center}
\includegraphics[width=\linewidth]{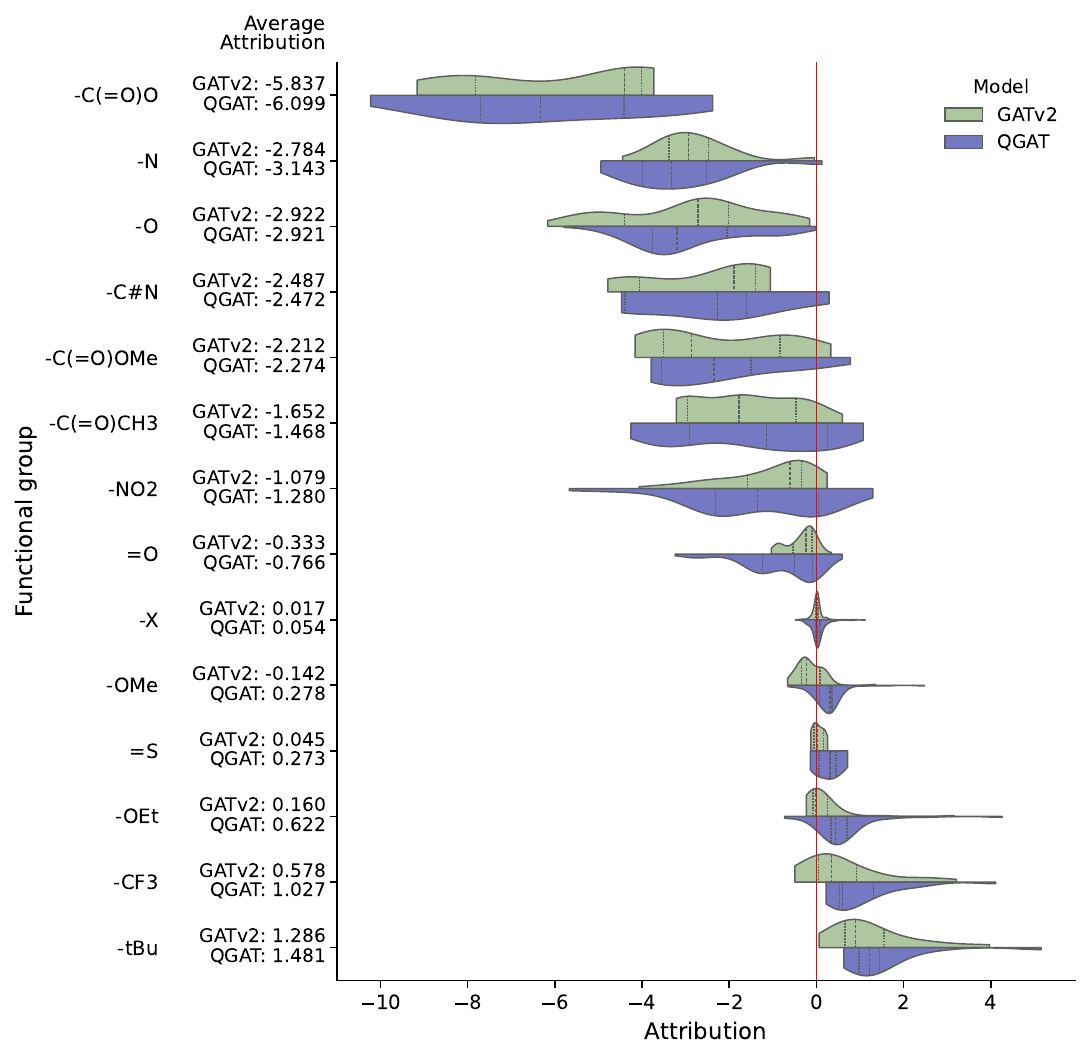}
\end{center}
\caption{The attributions of different functional groups in the whole FreeSolv dataset. Green colors represent GATv2 model, while blue colors represent QGAT model. }
\label{fig: SME_FreeSolv}
\end{figure}

\begin{figure}[h]
\begin{center}
\includegraphics[width=\linewidth]{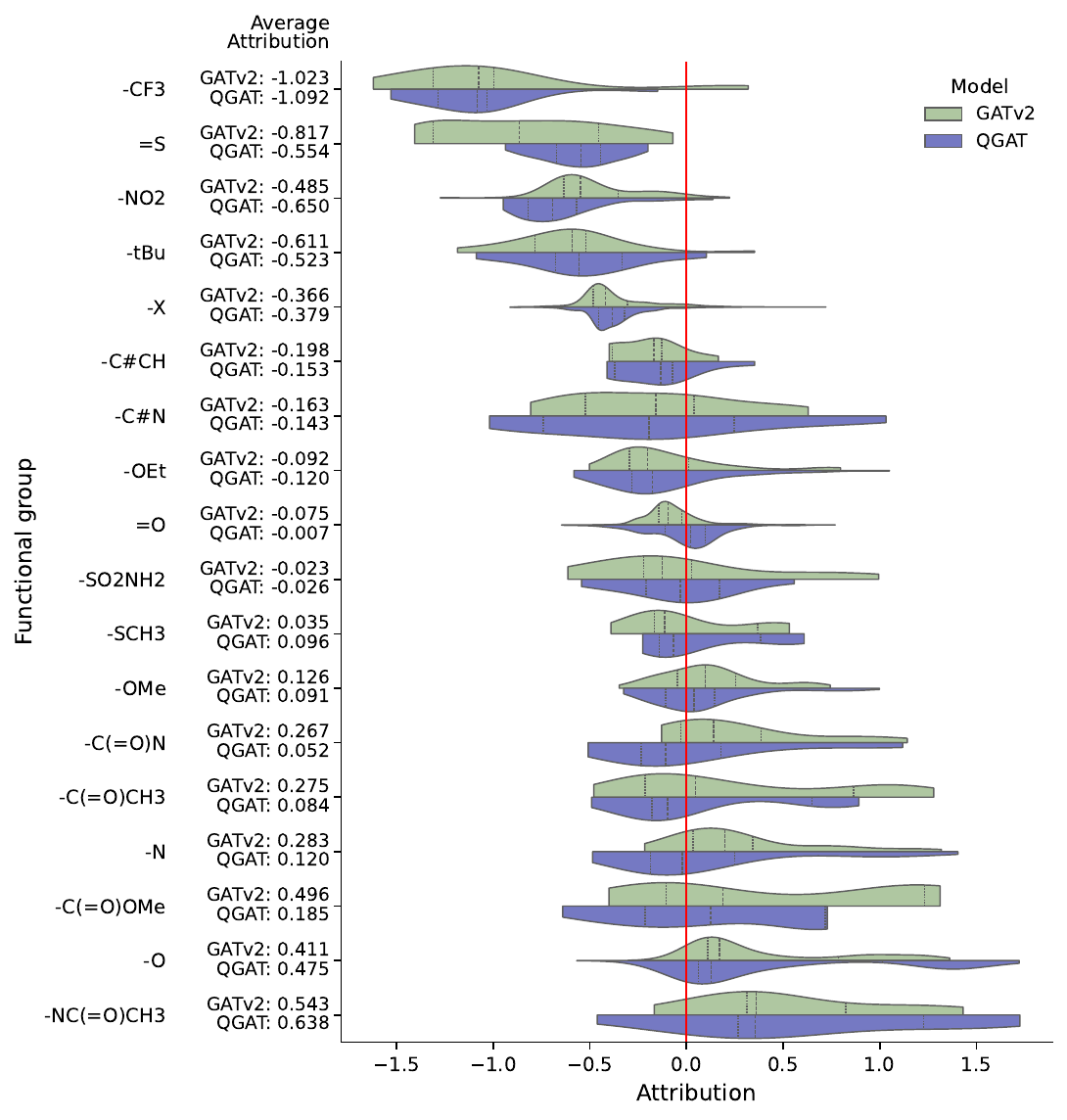}
\end{center}
\caption{The attributions of different functional groups in the whole ESOL dataset. Green colors represent GATv2 model, while blue colors represent QGAT model. }
\label{fig: SME_ESOL}
\end{figure}

\begin{figure}[h]
\begin{center}
\includegraphics[width=\linewidth]{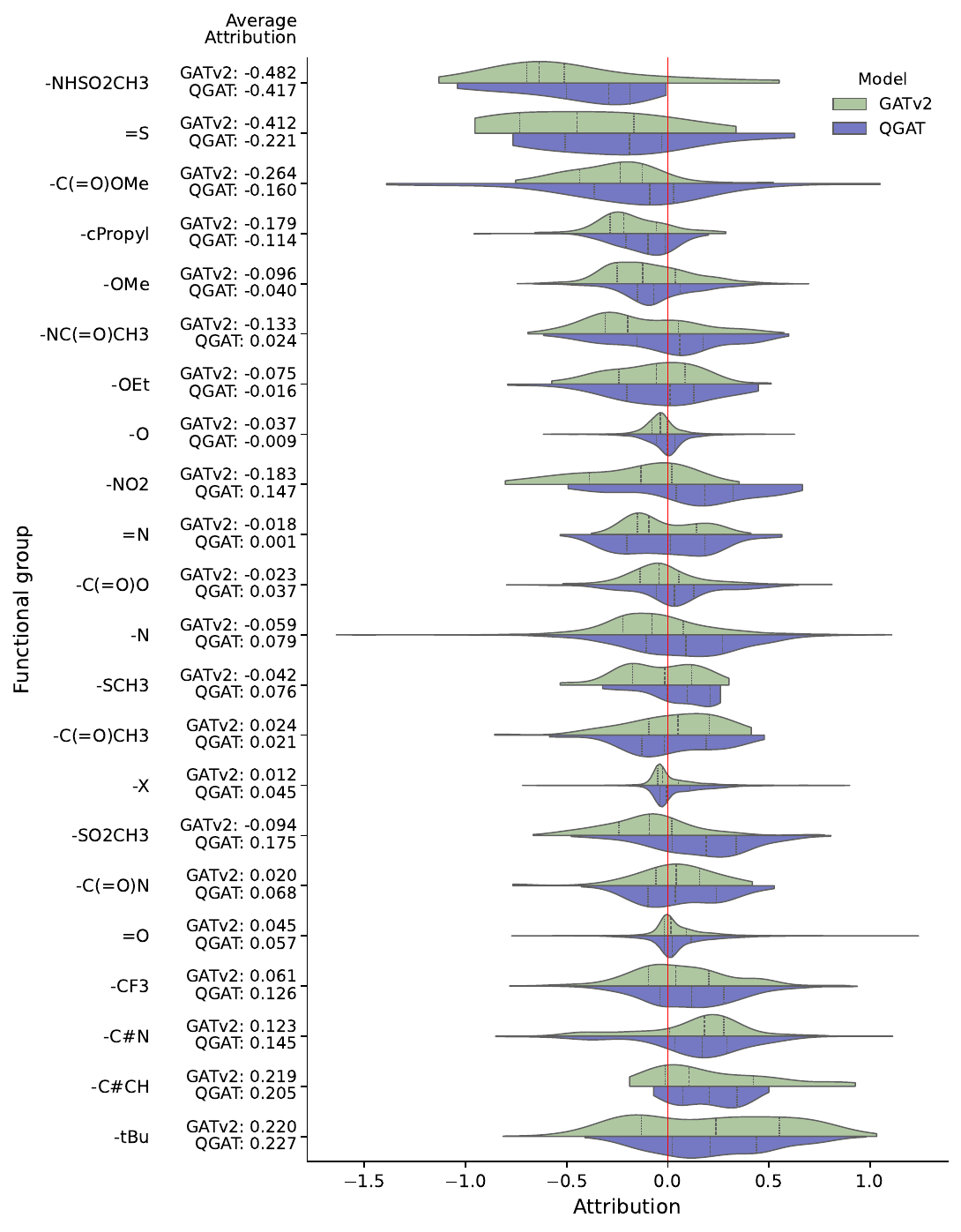}
\end{center}
\caption{The attributions of different functional groups in the whole ChEMBL-BACE dataset. Green colors represent GATv2 model, while blue colors represent QGAT model. }
\label{fig: SME_ChEMBL}
\end{figure}

Figures~\ref{fig: SME_BACE}--\ref{fig: SME_ChEMBL} summarize the full attribution distributions for GATv2 and QGAT. 
Across datasets, many functional groups show similar attribution directions between the two models, while differences are more apparent in attribution magnitude and distributional spread for selected groups. 
Several functional groups also exhibit broad distributions spanning both positive and negative values, indicating that their contributions are context dependent rather than determined solely by functional-group identity. These distributions are therefore used to characterize model specific attribution patterns across molecular contexts, rather than to assign universal favorable or unfavorable effects to individual chemical groups.

\section{Additional Analysis of External Medicinal Chemistry Series}
\label{app:MK8931}

\subsection{Integrated Gradients for the Verubecestat Series}
\label{app:ig}
Integrated Gradients (IG) \citep{Sundararajan2017IG, Kuang2025IG, Wang2025IG} is a gradient-based attribution method that quantifies the contribution of input features to a model prediction relative to a reference input. Rather than evaluating the gradient at the observed input, IG accumulates gradients along a straight line path from a baseline $x'$ to the input $x$.  

Let $F(x)$ denote the model output of interest. The IG of the $i$-th input feature is defined as

\begin{equation}
\mathrm{IG}_i(x)=(x_i-x_i')\int^1_{\alpha=0} 
\frac{\partial F (x'+\alpha (x-x'))}{\partial x_i}d\alpha
\label{eq:IG}
\end{equation}

where $x'$ denotes the baseline input and $\alpha$ parameterizes the interpolation path between $x'$ and $x$. The resulting attribution reflects the accumulated sensitivity of the model output to feature $i$ along this path.

\subsection{Activity Cliff  Analysis for the Verubecestat Series}
\label{app:activity_cliff}

We provide the complete integrated gradient visualizations for the two AC pairs identified in the Verubecestat series, compounds \textbf{21}--\textbf{24} and \textbf{30}--\textbf{32}. As described in the main text, these pairs satisfy the AC criterion of a potency difference $|\Delta pK_i|>1.0$ between structurally similar analogues.

\begin{figure}[t]
\centering
\includegraphics[width=\linewidth]{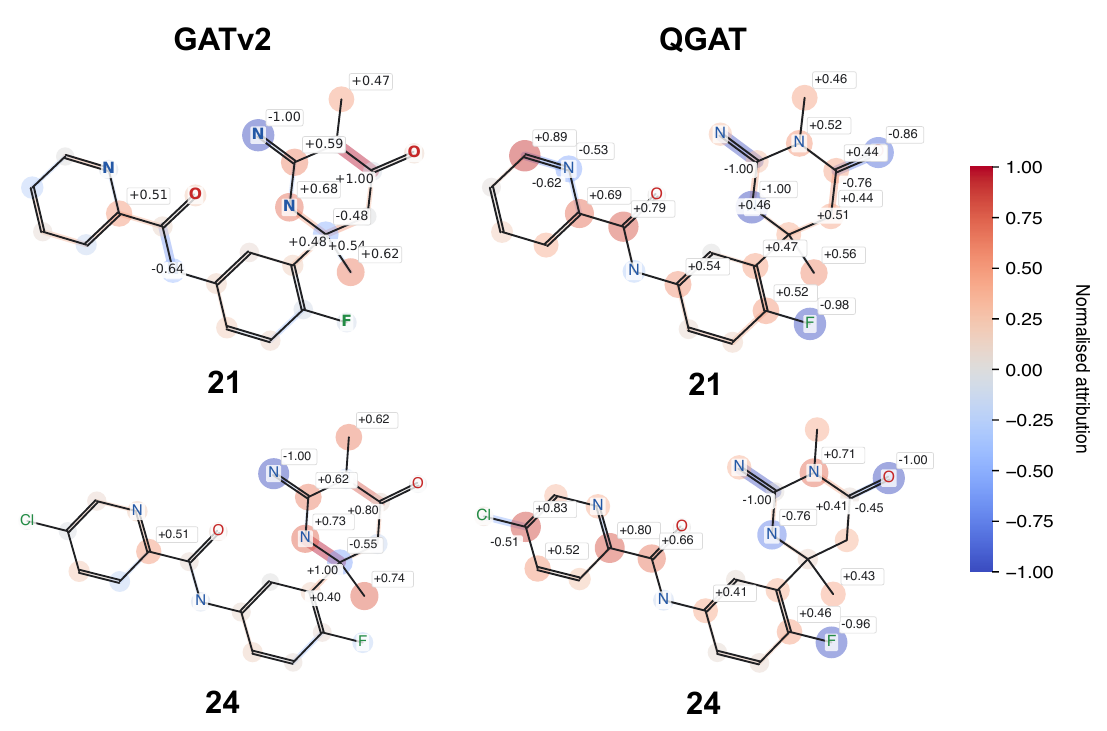}
\caption{IG attribution maps for the AC pairs identified in the Verubecestat series: compounds \textbf{21}--\textbf{24}. 
Attributions are normalized within each molecule for visualization. 
The figure provides the complete atom level attribution patterns corresponding to the qualitative AC analysis discussed in the main text.}
\label{fig:AC_21_24}
\end{figure}

\begin{figure}[h]
\centering
\includegraphics[width=\linewidth]{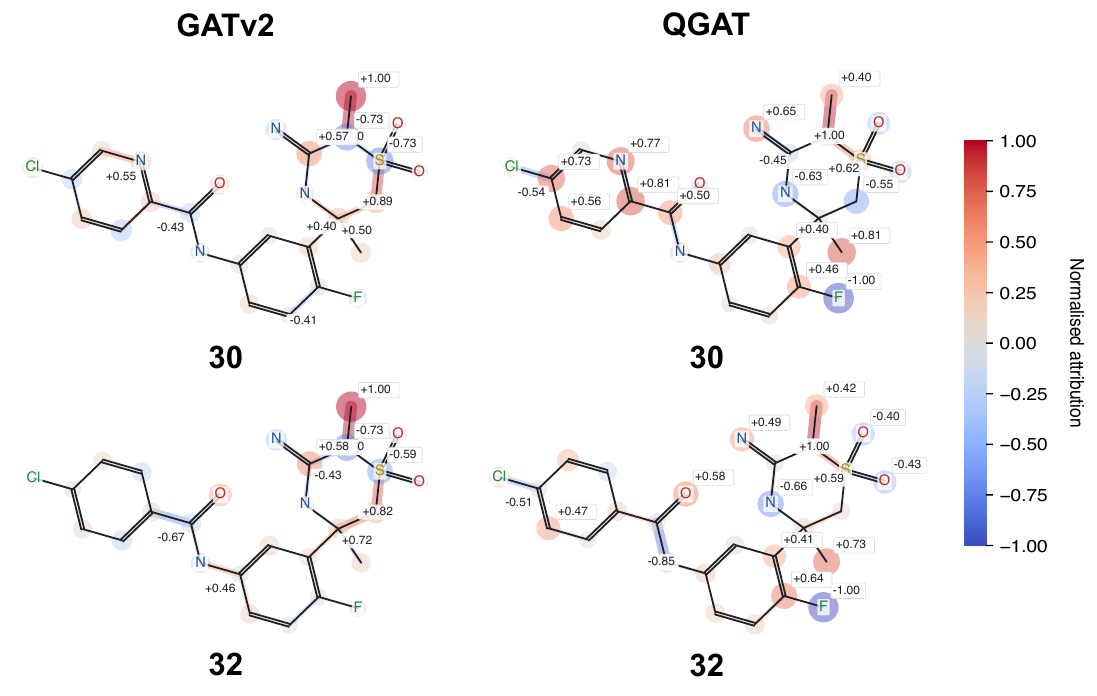}
\caption{IG attribution maps for the AC pairs identified in the Verubecestat series: compounds \textbf{30}--\textbf{32}. 
Attributions are normalized within each molecule for visualization. 
The figure provides the complete atom level attribution patterns corresponding to the qualitative AC analysis discussed in the main text.}
\label{fig:AC_30_32}
\end{figure}

Figures~\ref{fig:AC_21_24} and ~\ref{fig:AC_30_32} show the IG attribution maps for both GATv2 and QGAT.  The visualization is intended to complement the main text analysis by showing the full spatial redistribution of attribution induced by the corresponding structural modifications.  Because only two AC pairs are available in this series, we treat these examples as qualitative case studies rather than as a statistical evaluation of AC prediction.

\subsection{Additional External Series: Umibecestat and LY3202626}
\label{app:additional}
We further evaluate two independent medicinal chemistry case studies, Umibecestat (CNP520) \citep{CNP5202021} and LY3202626 \citep{LY3202626}. Because these sets are small and contain several compounds with high similarity to the training data, we do not treat them as chemical OOD benchmarks. Instead, they are used to assess whether the ranking behavior observed for Verubecestat generalizes qualitatively across additional analogue series.

For the Umibecestat series, QGAT yields a modestly higher mean Spearman correlation than GATv2 (0.692 vs. 0.655), while RMSE remains comparable. For the LY3202626 series, the two models achieve nearly identical mean Spearman correlation (0.788 vs. 0.784), whereas GATv2 obtains lower RMSE.

\end{document}

%% file: macros.tex
\definecolor{lightyellow}{rgb}{1.0, 0.95, 0.7}
\definecolor{Blue}{rgb}{0, 0, 0.8}
\definecolor{blue}{rgb}{0,0,1}
\definecolor{mydarkblue}{rgb}{0,0.08,0.45}
\definecolor{mydarkblue2}{rgb}{0.133, 0.133, 0.698}
\definecolor{echodrk}{HTML}{0099cc}
\definecolor{mymauve}{rgb}{0.58,0,0.82}
\definecolor{darkgreen}{rgb}{0,0.40,0}
\definecolor{firebrick}{rgb}{0.698,0.133,0.133}
\definecolor{midnightblue}{rgb}{0.1,0.1,0.44}
\definecolor{citeblue}{RGB}{0, 113, 188}
\definecolor{oxfordblue}{rgb}{0.0,0.13,0.28}
\definecolor{prussianblue}{rgb}{0.0,0.19,0.33}
\definecolor{coolteal}{rgb}{0, 0.45, 0.45}
\definecolor{olive}{rgb}{0.1, 0.3, 0}
\definecolor{mypurple}{rgb}{0.5,0,0.5}
\definecolor{almond}{rgb}{0.94, 0.87, 0.8}

\definecolor{blue_ampEncoding}{HTML}{DAE8FC}
\definecolor{green_encoder}{HTML}{D5E8D4}
\definecolor{purple_decoder}{HTML}{E1D5E7}
\definecolor{yellow_measure}{HTML}{FFF2CC}
\definecolor{gray_block}{HTML}{F5F5F5}
\definecolor{pink_dru}{HTML}{FAD9D5}
\definecolor{orange_v}{HTML}{FAD7AC}
\definecolor{Lightblue}{HTML}{E7F4FC}

\DeclareDocumentCommand \norm { o m }{{\lVert #2 \rVert_#1}}

\declaretheoremstyle[%
  spaceabove=10pt,%
  spacebelow=2pt,%
  headfont=\normalfont\itshape,%
  postheadspace=0em,%
  qed=%
]{prfstyle}